\documentclass[a4paper]{article}

\usepackage{graphicx}
\usepackage{amsfonts}
\usepackage{amsmath}
\usepackage{amsbsy}
\usepackage{subfigure}
\usepackage{geometry}
\usepackage{setspace} 
\usepackage{bm}
\usepackage{multicol}
\usepackage{multirow}
\usepackage{color} 

\title{  
Dynamics of braided coronal loops -- I.  Onset of magnetic reconnection}
\author{A.L.~Wilmot-Smith, D.I.~Pontin and G.~Hornig. \\
{\small Division of Mathematics, University of Dundee, Dundee, DD1 4HN, UK}}

\begin{document}
\maketitle

\abstract{}
{The response of the solar coronal magnetic field to small-scale photospheric boundary motions 
including the possible formation of current sheets via the Parker scenario is one of open questions of 
solar physics. Here we address the problem via a numerical simulation.}
{The three-dimensional evolution of a braided magnetic field which is initially close to a force-free state 
is followed using a resistive MHD code.}
{A long-wavelength instability takes place and leads to the formation of two thin current
layers.   Magnetic reconnection occurs across the current sheets with three-dimensional features shown, 
including an elliptic magnetic field structure about the reconnection site,
and results in an untwisting of the global field structure.}

\vspace*{0.5cm}

\newpage

\section{ Introduction}
\label{sec:introduction}

Parker's notion of the `topological dissipation' of coronal magnetic fields (Parker, 1972) continues
to generate much debate.  Simply put, Parker's suggestion is that following boundary motions of sufficient 
complexity the magnetic field of a coronal loop will be unable to ideally relax to a smooth force-free
equilibrium and instead tangential discontinuities in the field, corresponding to current sheets, 
will develop.   
In general terms the possible outcomes of relaxation are a development of singular current sheets
(e.g. Ng \& Bhattacharjee, 1998; Janse \& Low, 2009), development of thin but non-singular current layers
(e.g. Longcope \& Strauss, 1994;  Galsgaard {\it et al.}~2003), and a  smooth equilibrium with large-scale current 
features (e.g. van Ballegooijen 1985; Craig \& Sneyd, 2005, Wilmot-Smith {\it et al.}~2009a).
It should also be noted 
 that the distinction between the first two cases may be difficult to determine numerically.

\begin{figure}[tb]
\begin{center}
\includegraphics[width=0.36\textwidth]{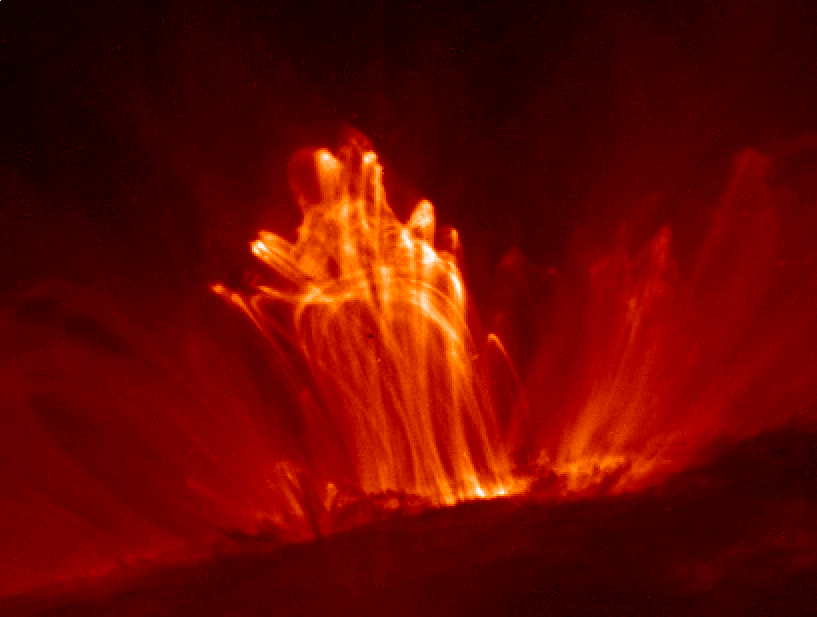}
\includegraphics[width=0.35\textwidth]{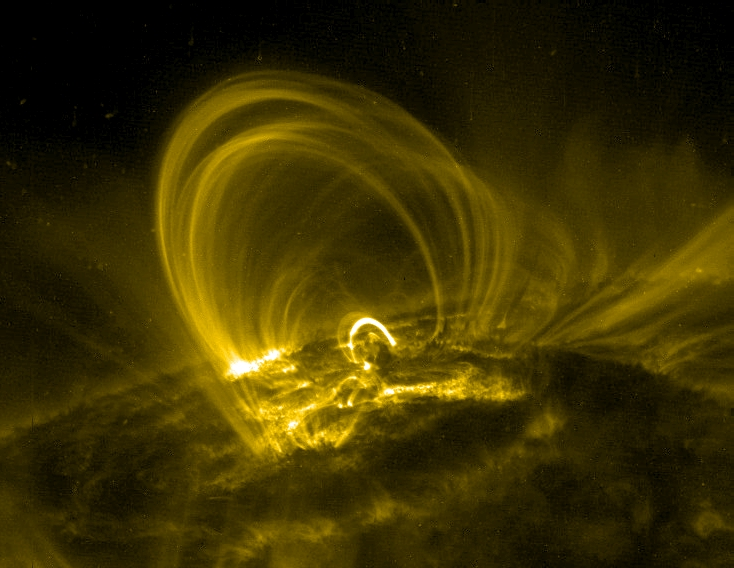}
\end{center}
\caption{
TRACE coronal loops. ({\it Left}) A large-scale tangled configuration 
and ({\it right)} apparently smooth loops.}
\label{fig:traceloops}
\end{figure}

In a previous a paper (Wilmot-Smith {\it et al.}~2009a) we {  considered
 the ideal relaxation  of a braided magnetic field towards a force-free equilibrium.
The magnetic field configuration was based on the pigtail braid and}
 imposed as an initial condition (rather than being built up through boundary motions).
 {    This particular braid was chosen since it is the simplest non-trivial braid with no net twist
and, accordingly, is the most conservative realistic case to examine.   More complex braided fields will, 
in general, contain components of the pigtail-type.  
There is ample motivation for modelling loops as having braided components.
Photospheric turbulence subjects loop footpoints to a random walk,
with motions of the fragments about each other acting to braid (or unbraid) the overlying loop.
However, while there is some evidence for the existence of braided loop configurations
(see the left-hand image of Figure~\ref{fig:traceloops} for example), most coronal
loops appear to be close to a potential field (e.g.~Figure~\ref{fig:traceloops}, right-hand image).
This may be an effect of the large aspect ratios typical to loops, since a winding of a field line
around another field line is almost undetectable when smoothed out over the length of the loop
(see also Berger \& Asgari-Targhi, 2009).  Another reason could be that reconnection is very 
efficient in maintaining a low degree of topological complexity in loops.  The present work is designed 
in part to test whether this is the case.}

In Wilmot-Smith {\it et al.}~(2009a) an ideal Lagrangian relaxation scheme (Craig \& Sneyd 1986; Pontin {\it et al.} 2009) 
was used to ideally evolve {  the braided} field towards a force-free state.
A smooth near force-free equilibrium was attained with large-scale current features
i.e. without any tangential discontinuities or even strong current concentrations.
(This situation may be compared with 
Parker (1994) where the same pigtail braid situation is 
considered as a thought experiment and concluded to inevitably lead to tangential discontinuities.
It appears that Parker's assumption of $\alpha = 0$ in the domain fails; we indeed have $\int \alpha \ dS=0$, 
where $S$ is a cross-sectional surface through the braid, but $\alpha$ varies significantly between field lines).

Although the local current density in the ideal equilibrium is of large-scale, a global quantity, the
integrated parallel current,
$\int J_{\|} \ dl$, has small scales (here the parallel indicates parallel to the magnetic field 
and the integral 
is taken along magnetic field lines).  By small scales we mean that  $\int J_{\|} \ dl$  varies significantly 
between neighbouring field lines as the field lines pass though a particular plane, such as the lower boundary.
In Wilmot-Smith {\it et al.}~2009a it  was suggested that these small scales in $\int J_{\|} \ dl$ could lead to the development of a resistive 
instability and so a loss of equilibrium of the field.  This mechanism would be distinct to that of Parker's
topological dissipation but have the same consequence.

The consideration that for sufficiently small scales in the integrated parallel current then 
a force-free magnetic field will be resistively unstable 
(the instability being dependent, for a given scale in the integrated parallel current, on the value of the resistivity)  
motivated us to put the end, near force-free, state of the Lagrangian relaxation into a resistive MHD code and 
test whether or not the state is stable (for various values of resistivity allowed by numerical limitations).  
It turns out that the braided field {  as implemented in the resistive code is not stable.  This finding,}
 together with the subsequent evolution of the field, allows us to address a number of key questions related to MHD 
 and the behaviour of  coronal magnetic fields.  The results are described in this series of papers.  Here we describe the details 
relating to the numerical setup of the problem {  and the early evolution of the system.}
A subsequent paper addresses the long term evolution of the system.

\section{Numerical Scheme and Simulation Setup}
\label{sec:numerical}

\subsection{Numerical Scheme}
\label{sec:code}

The numerical scheme employed in the simulations that follow is described
briefly below (further details may be found in Nordlund \& Galsgaard (1997) and at
http://www.astro.ku.dk/$\sim$kg). We solve the three-dimensional resistive MHD
equations in the form
\begin{eqnarray}
\frac{\partial {\bf B}}{\partial t} & = & - \nabla \times {\bf E},
\label{numeq1}\\ 
{\bf E} & = & -\left( {\bf v} \times {\bf B} \right)
\: + \: \eta {\bf J}, \label{numeq2}\\ 
{\bf J} & = & \nabla \times
{\bf B}, \label{numeq3}\\ 
\frac{\partial \rho}{\partial t} & = & -
\nabla \cdot \left( \rho {\bf v} \right), \label{numeq4}\\
\frac{\partial}{\partial t}\left( \rho {\bf v} \right) & = & - \nabla
\cdot \left( \rho {\bf v} {\bf v} \: + \: {\underline {\underline
\tau}} \right) \: - \: \nabla P \: + \: {\bf J} \times {\bf B},
\label{numeq5}\\ 
\frac{\partial e}{\partial t} & = & -\nabla \cdot
\left( e {\bf v} \right) \: - \: P \: \nabla \cdot {\bf v} \: + \:
Q_{visc} \: + \: Q_{J} \label{numeq6},
\end{eqnarray}
where ${\bf B}$ is the magnetic field, ${\bf E}$ the electric field,
${\bf v}$ the plasma velocity, $\eta$ the resistivity, ${\bf J}$ the
electric current, ${\rho}$ the density, ${\underline {\underline
\tau}}$ the viscous stress tensor, $P$ the pressure, $e$ the internal
energy, $Q_{visc}$ the viscous dissipation and $Q_{J}$ the Joule
dissipation. An ideal gas is assumed, and hence $P \: = \: \left(
\gamma -1 \right) \: e \: = \: {\textstyle \frac{2}{3}}e$.
These equations have been non-dimensionalised by setting
the magnetic permeability $\mu_0 = 1$, and the gas constant ($\mathcal{R}$)
equal to
the mean molecular weight ($M$). The result is that for a volume in which $| \rho |=| {\bf B} | = 1$, time 
units are such that an Alfv\'{e}n wave would travel one space unit in one unit of time. 

The equations (1-6) are solved on staggered meshes; with respect to a mesh on which $\rho$ and $e$ 
are defined at the body centre of the cell, $\mathbf{B}$ and $P$ are defined at face centres and
 $\mathbf{E}$ and $\mathbf{J}$ at edge centres.  In this way the required MHD
conservation laws are automatically satisfied.  Derivatives are calculated using sixth-order finite 
differences that return a value which is shifted half a grid-point up or down with respect to the input 
values. When the staggered mesh means that some quantity must be interpolated,  data values are 
calculated using a fifth-order interpolation method at the relevant position.  A third-order 
predictor-corrector method is employed for time-stepping. 

In all simulation runs we employ a spatially uniform resistivity model. Viscosity is calculated using a 
combined second-order and fourth-order method (sometimes termed `hyper-viscosity'), which is 
capable of providing sufficient localised dissipation where necessary to handle the development of 
numerical instabilities (Nordlund \& Galsgaard 1997). The effect is to `switch on' the viscosity where 
very short length scales develop, while maintaining a minimal amount of viscous dissipation where 
the velocity field is smooth. 

\subsection{Creating the Initial Condition}
\label{sec:ICinterp}

As discussed in Sec.~\ref{sec:introduction}, the initial state for the resistive MHD simulations is drawn 
from the final state of the Lagrangian relaxation experiment of Wilmot-Smith {\it et al.}~2009a.
The quantities previously known from the Lagrangian code (see Craig \& Sneyd 1986)
 are the magnetic field ${\mathbf{B}}$ and the current ${\mathbf{J}}$ and 
  in the near force-free relaxed state these are known on a highly distorted mesh.
We describe below the process of constructing the field on the rectangular grid required for the
 resistive MHD simulations.

In order to ensure that the initial magnetic field 
is divergence-free,  we work first with the vector potential $\mathbf{A}$ for $\mathbf{B}$.
In the relaxation scheme, the calculation of $\mathbf{A}$ requires only that we know the 
initial vector potential before relaxation and the mesh deformation. 
In terms of the initial mesh 
$\mathbf{X}$ and the final `relaxed mesh' $\mathbf{x}$, the $i$th component of the final vector potential 
$\mathbf{A}^f$ is given (see Appendix) in terms of the initial vector potential $\mathbf{A}^0$ by
\begin{equation}
\label{eq:Atoexplain}
A^f_i = \sum_{j=1}^3 A^0_j \frac{{\partial X_j}}{\partial x_i}.
\end{equation}

To create the input magnetic field for the MHD simulations we then interpolate this vector potential onto 
a rectangular grid. Since the magnetic field components are face-centred on the staggered grid, the 
vector potential components are interpolated onto locations corresponding to edge centres of the 
desired grid. We then obtain the magnetic field by taking the $curl$ of $\mathbf{A}$ using the sixth-order 
finite differences described above, which yields magnetic field components at face-centres as required. 
An interpolation scheme using biharmonic spline radial basis functions was applied to 
$\mathbf{A}$, the particular scheme chosen to maximize the smoothness
 of the corresponding current density $\mathbf{J}$, which involves second derivatives of $\mathbf{A}$. 
 To further improve this smoothness a simple five-point smoothing algorithm was finally applied to 
 $\mathbf{A}$, before taking the $curl$.

The result of the above is that the initial braided magnetic field for our MHD simulations is divergence-free 
to accuracies on the order of truncation errors of the sixth-order finite differences (with typical maximum 
$|\nabla \cdot {\bf B}| \sim 10^{-6}$ within the domain).   The topology of the magnetic field turns out to be 
well conserved by the process, another important consideration for the experiment.
However, a drawback is that the quality of the force-free approximation is not perfectly
maintained; the initial state is further from force-free than the relaxed field of the Lagrangian 
experiment.   Details and implications are discussed in Secs.~\ref{sec:results} and \ref {sec:discussion}.

\subsection{Initial State}

\begin{figure}[htb]
\begin{center}
\includegraphics[width=0.236\textwidth]{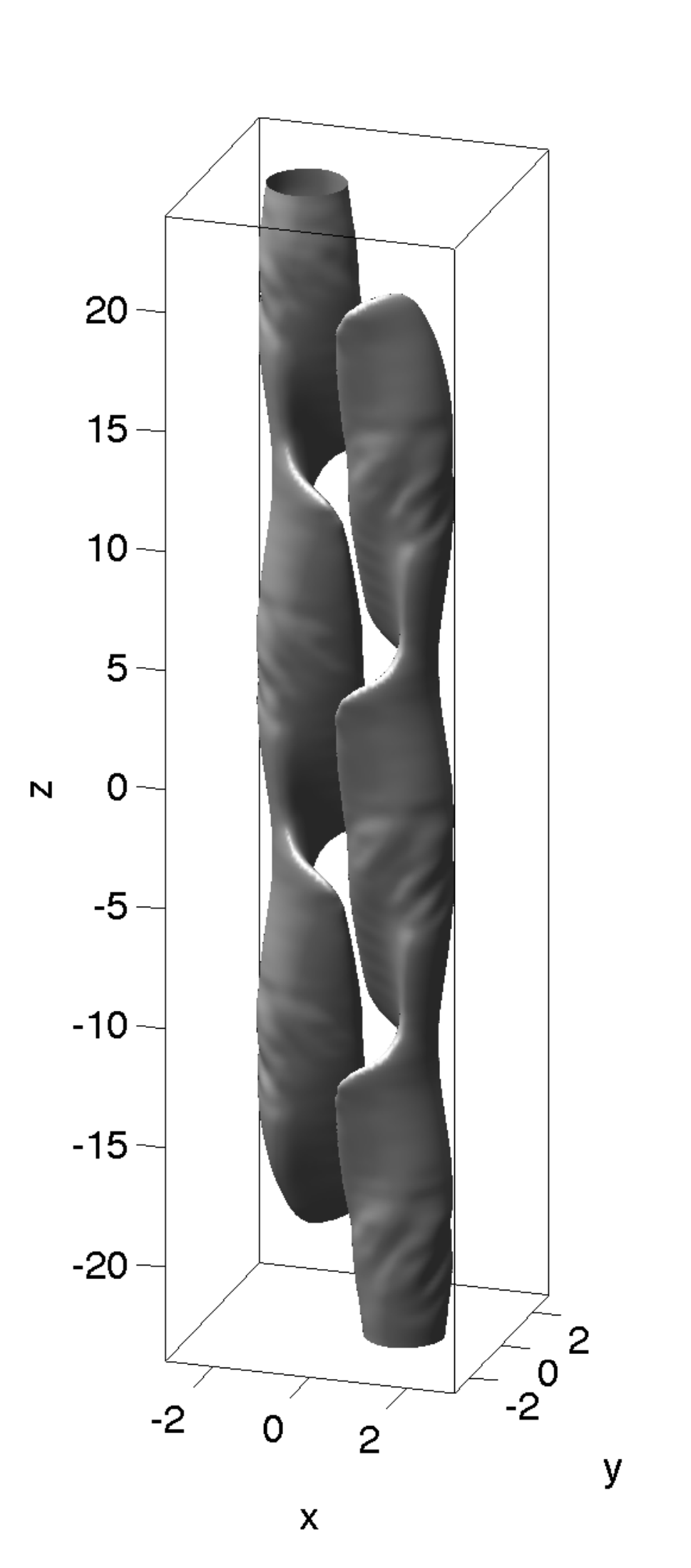}
\includegraphics[width=0.241\textwidth]{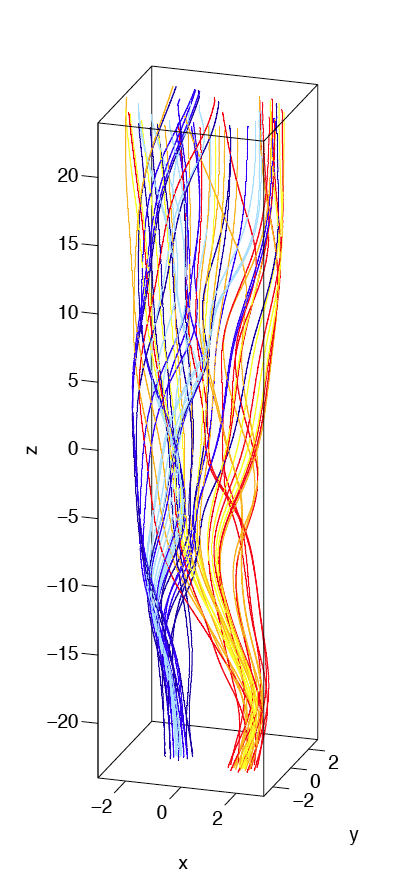}
\end{center}
\caption{Initial state of the simulation:
({\it left}) Isosurface of current $\vert \mathbf{J} \vert$ at 25\% of the domain maximum
and ({\it right)} some particular magnetic field lines illustrating the braided
nature of the field. }
\label{fig:initialstate}
\end{figure}

The initial magnetic field is given on a domain of size $[-24,24]$ in the vertical ($z$) direction and 
$[-6,6]$ in both the horizontal ($x,y$) 
directions.  Covering this domain we take a uniform mesh of $320^{3}$ cells and employ
closed boundary conditions  in all three directions.
The magnetic field is line-tied, and can be very closely approximated by $\mathbf{B}=[0,0,1]$ at 
the boundaries, i.e.~it is directed perpendicular to the $z$ boundaries and parallel to the $x$ 
and $y$ boundaries. 
To achieve the perpendicular condition the Lagrangian relaxation experiment was re-run on the 
larger horizontal domain (now $[-6,6]^{2}$ compared with $[-4,4]^{2}$ in Wilmot-Smith {\it et al.}~2009a).
The braided field is centered in the middle of the domain and the field is close to uniform in the
region external to the braid.  Accordingly the results presented here are shown for only the subsection  
of the full domain in which the important dynamics occur, specifically $[-3,3]^2 \times [-24,24]$.

An isosurface of current in the initial state is given the left-hand image of Fig.~\ref{fig:initialstate}.
The current has large-scales  (see also the upper-left hand image of Fig.~\ref{fig:jevolution} where 
contours of current in a horizontal cross-section are shown) with two fingers of current extending 
vertically through the domain.  Some sample field lines further illustrating the nature of the field are 
given in the right-hand image of the same figure.  
{Although non-trivial, the initial state has little magnetic energy in excess of the homogeneous field
($0.96\%$ in $[-3,3]^2 \times [-24,24]$).
The aspect ratio employed in the model is 1:8.  Although this is larger than that of many previous
simulations it is smaller than a realistic ratio for a coronal loop (1:50, say).  
In the configuration the poloidal field components are small compared with the toroidal 
components so that the field lines look almost straight.  This level of braiding would be observationally
difficult to distinguish from a potential field.}

{  To initialise the simulation
the dimensionless plasma density  (Sec.~\ref{sec:code}) is set at $\rho=1$ 
throughout the domain and the internal energy as $e=0.1$.}
The result is a plasma-$\beta$ at $t=0$ that lies in the range $\beta \in [0.1,0.14]$.  
For the results described in the main section of this paper (Sec.~\ref{sec:results}) we
consider the early evolution of the system (up to $t=14$) with time measured in 
units of the Alfv{\'e}n time.   A uniform resistivity of $\eta = 0.001$ has been taken and the effect of 
changing the resistivity is discussed at various points in the following text.

\section{Results}
\label{sec:results}

\subsection{Basic Properties}

\begin{figure}[htb]
\begin{center}
\includegraphics[width=7.1cm]{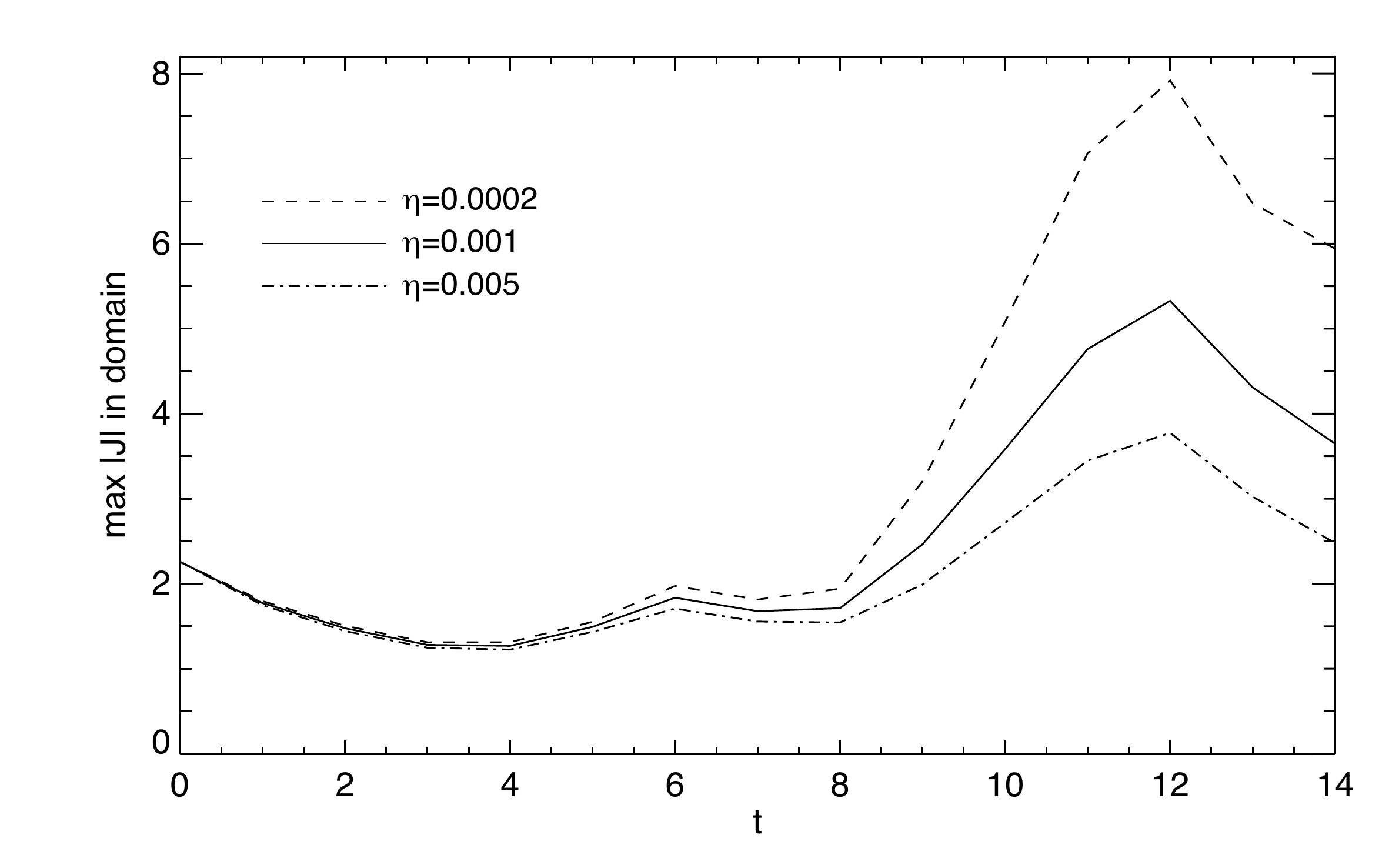}
\includegraphics[width=7.1cm]{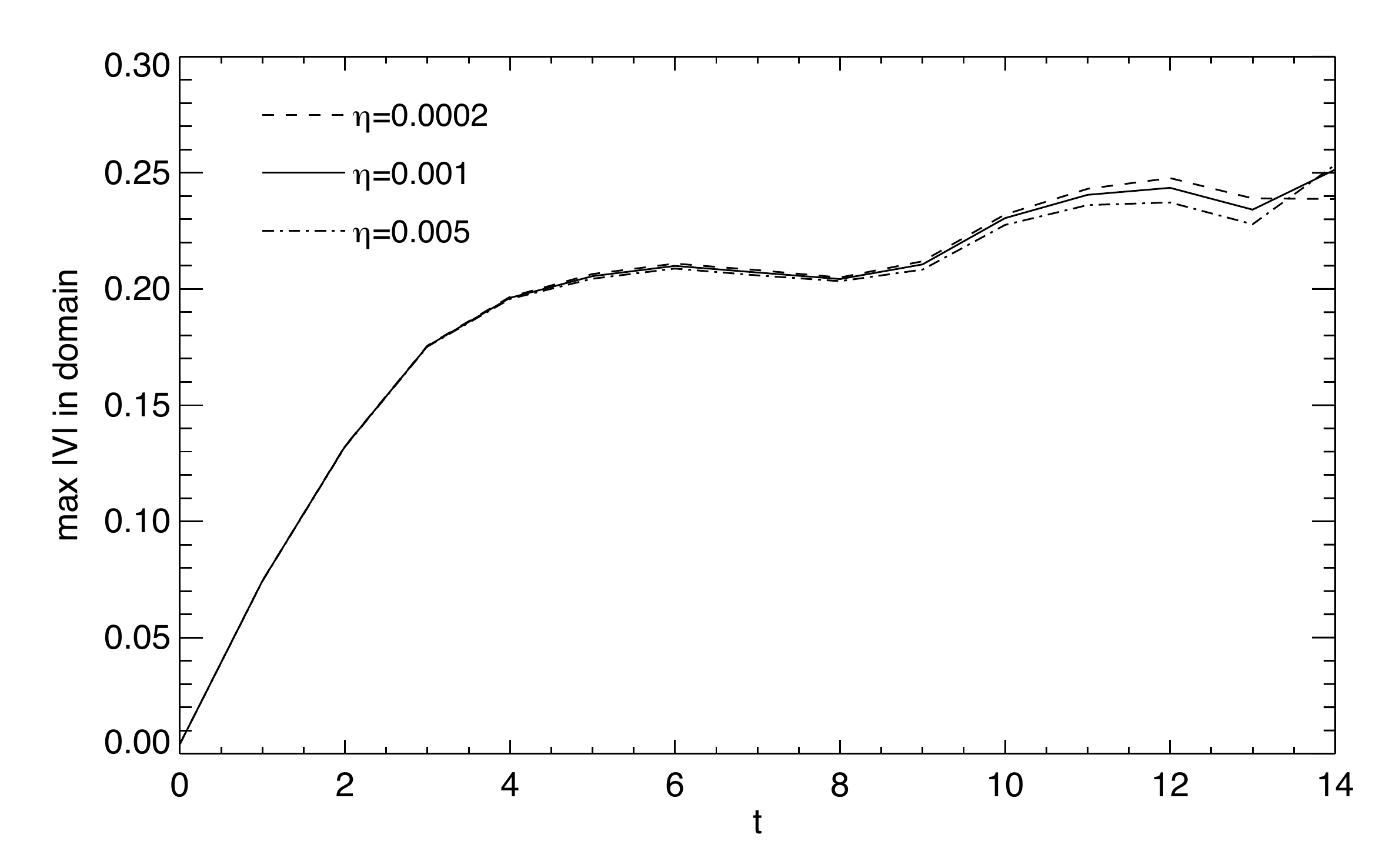}
\includegraphics[width=7.1cm]{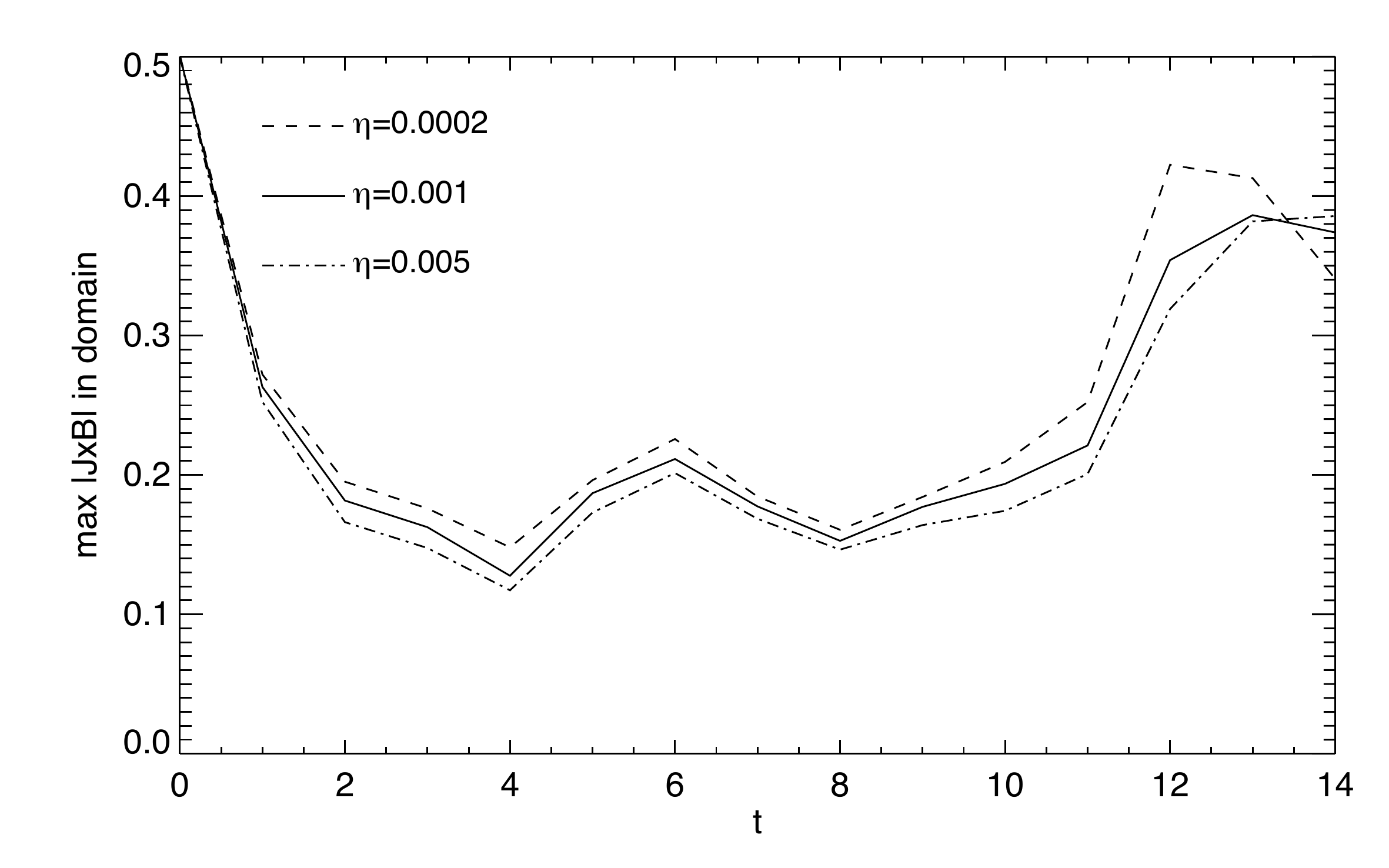}
\includegraphics[width=7.1cm]{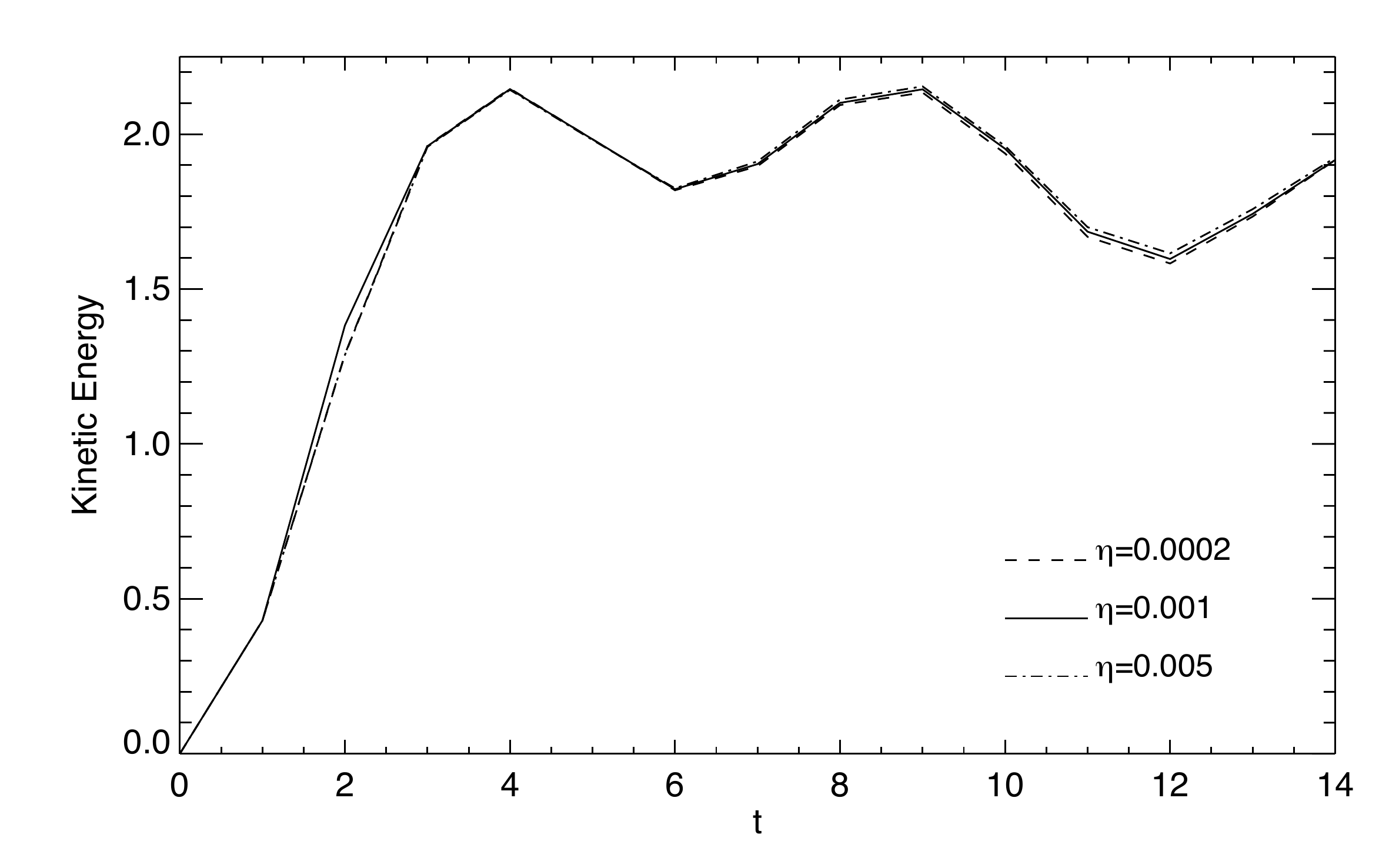}
\end{center}
\caption{Maximum absolute value of the current ({\it top left}), velocity ({\it top right}) and
Lorentz force ({\it bottom left}) and total kinetic energy ({\it bottom right}) in the domain with 
time for a sequence of decreasing uniform resistivies as indicated in the figures.}
\label{fig:jmax}
\end{figure}

Figure~\ref{fig:jmax} shows the maximum absolute values of the current, velocity and Lorentz force, 
and the kinetic energy ($\int \frac{1}{2} v^{2} dV$) for the time interval ($t \in [0,14]$) under 
consideration.  The domain taken in all cases is the central section, $[-3,3]^{2} \times [-24,24]$, of 
the full box. The variation in quantities is shown for a sequence of uniform resistivities decreasing
by over an order of magnitude, specifically $\eta = 0.005, 0.001$ and $\eta = 0.0002$.

The initially high maximum Lorentz force,  $\vert {\bf j} \times {\bf B} \vert_{max} =0.501$
decreases rapidly over the first few time units.  
Both the high value and the decrease are artifacts of the method used to create the initial state.  
The interpolation required to transfer the state to the Eulerian grid (as described in 
Sec.~\ref{sec:ICinterp}) results in some noise in the initial magnetic field and current density.
Some noise persists even with the application of a smoothing algorithm to the vector potential 
for the magnetic field and this is particularly noticeable in the Lorentz force rather than the magnetic 
field and current density alone.
Thus whilst the final state of the Lagrangian relaxation experiment had a very small 
maximum Lorentz force (specifically $\vert {\bf j} \times {\bf B} \vert_{max} \approx 2 \times 10^{-2}$),
the initial state here is further from force-free.   The decrease over $t \in [0,2]$ then arises though 
a smoothing of the noise in the initial state.

Turning now to the remaining quantities shown in Fig.~\ref{fig:jmax}, two primary  features are found.
Firstly a growth in the kinetic energy and maximum velocity for $t \in [0,4]$ occurs,
and this growth is independent of resistivity, $\eta$.
For the remaining time considered there is no significant change in kinetic energy.
{  Secondly, a linear growth in the current density occurs for $t \in [8,12]$.
The rate at which the maximum current density increases is higher for lower resistivity, $\eta$.   
At $t=12$ the maximum current density is achieved; this maximum is higher for lower resistivity
 but for all three resistivities the growth phase ends at the same time.

  The lack of dependence of kinetic energy on resistivity may suggest an ideal instability is present. 
The subsequent linear growth of current would then be
a non-linear consequence of this instability rather than its initial appearance.
This growth is clearly dependent on resistivity, being slower for higher values of $\eta$.
  Little is known about the non-linear  phase
of instabilities and such a dependence may still be consistent with an ideal instability
with a non-linear phase damped by $\eta$.
Strong conclusions are clearly difficult to draw at this stage.  An additional consideration is that the
implementation of the field on the new grid has resulted in significant Lorentz forces in the initial state.
We return to these questions in Section~\ref{sec:nature} but now proceed to consider the
 the nature of the currents within the domain, now fixing  $\eta = 0.001$.}

\subsection{Formation of Current Layers}

\begin{figure*}[]
\begin{center}
\includegraphics[width=0.194\textwidth]{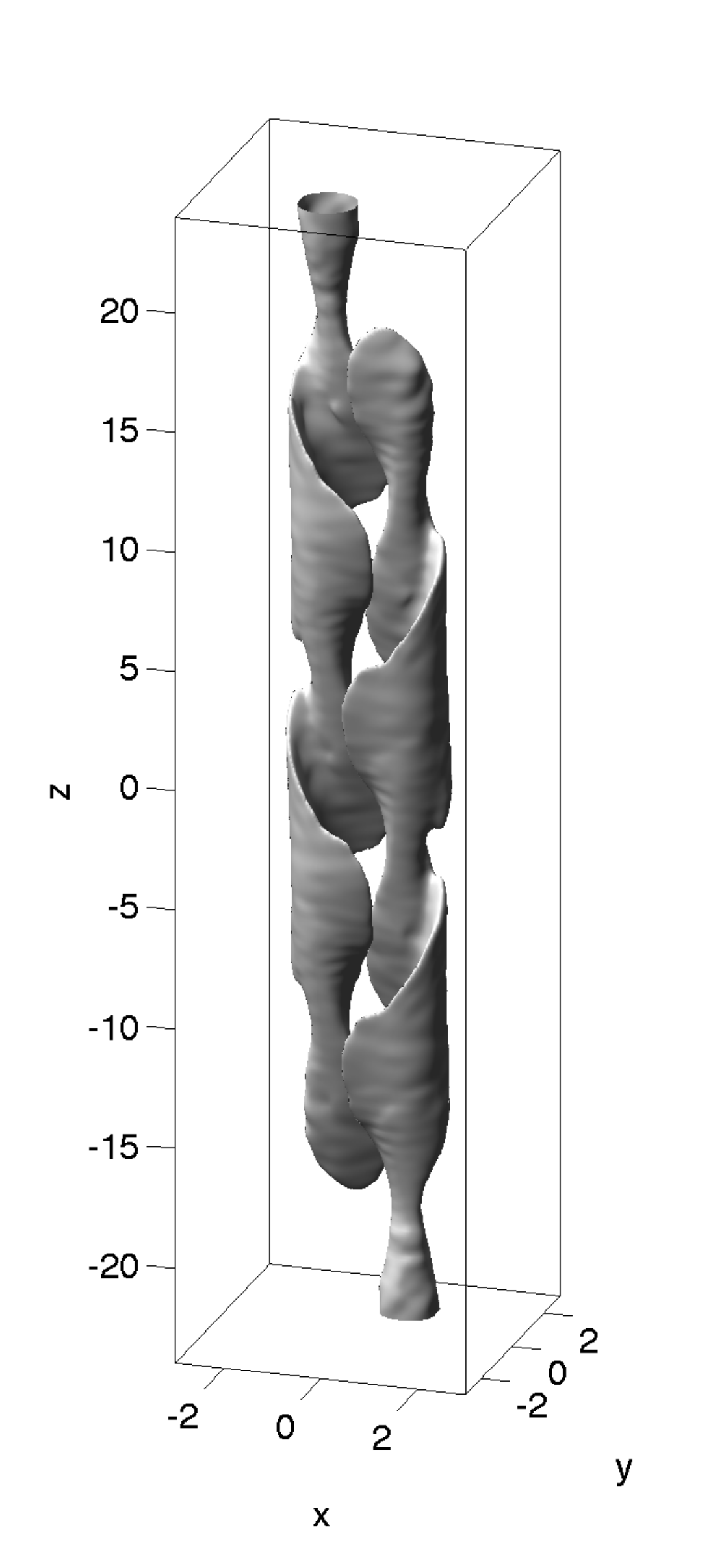}
\includegraphics[width=0.194\textwidth]{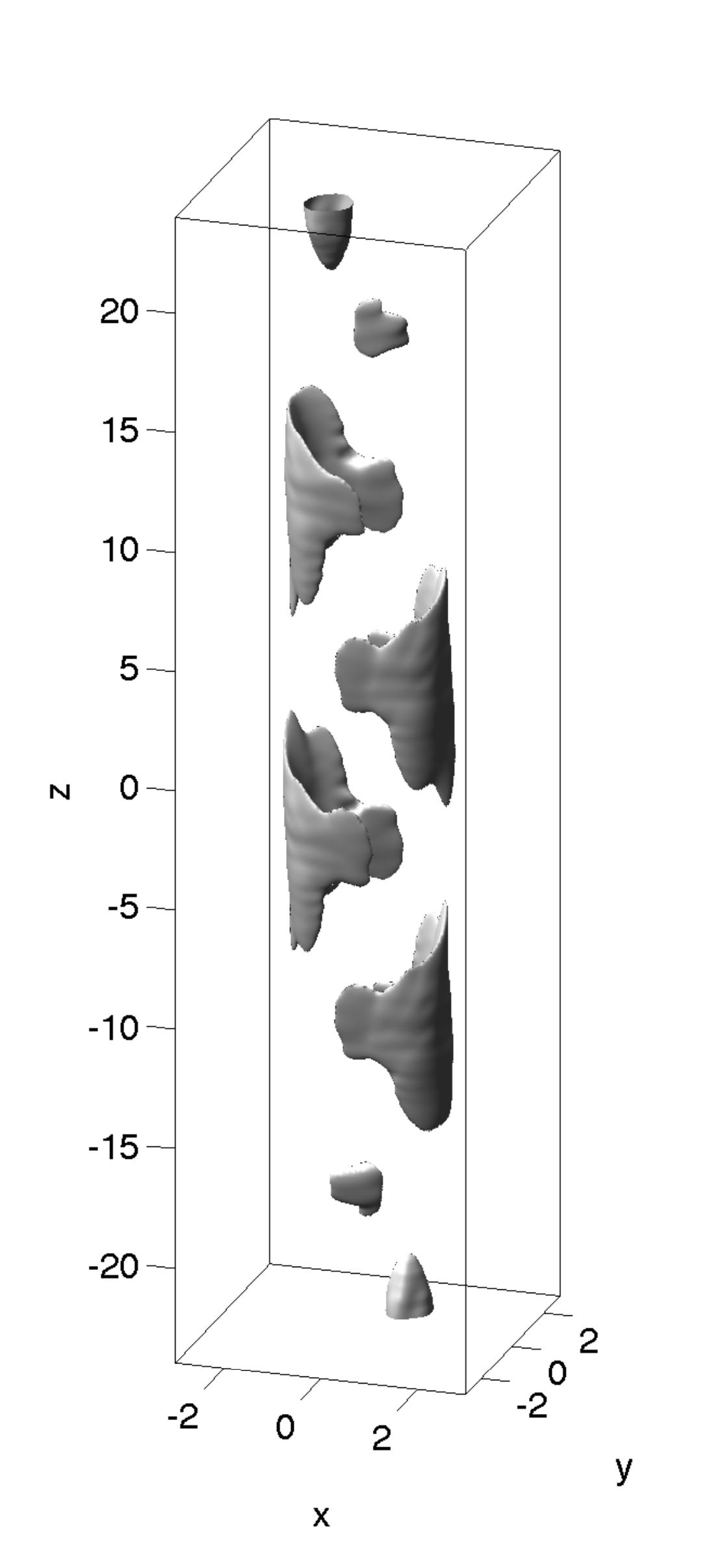}
\includegraphics[width=0.194\textwidth]{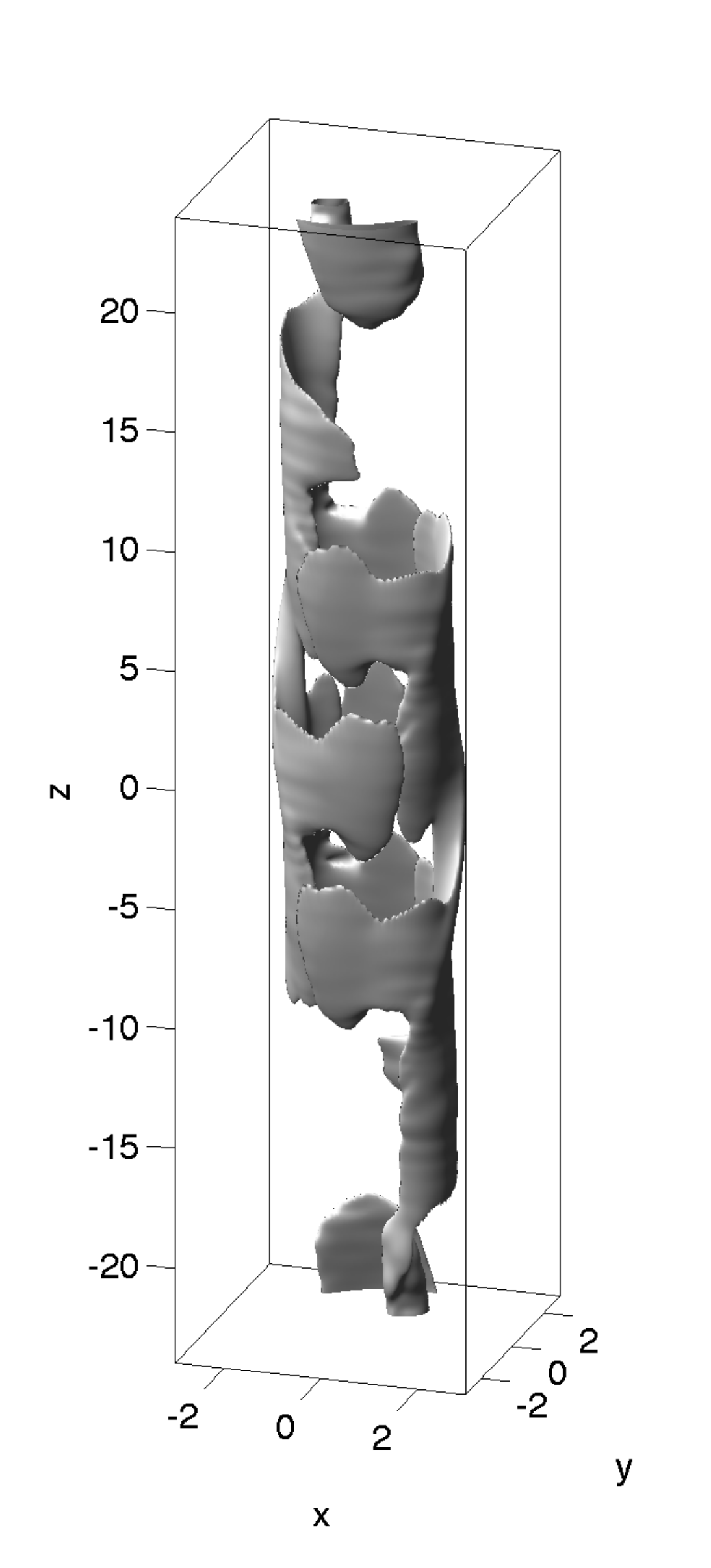}
\includegraphics[width=0.194\textwidth]{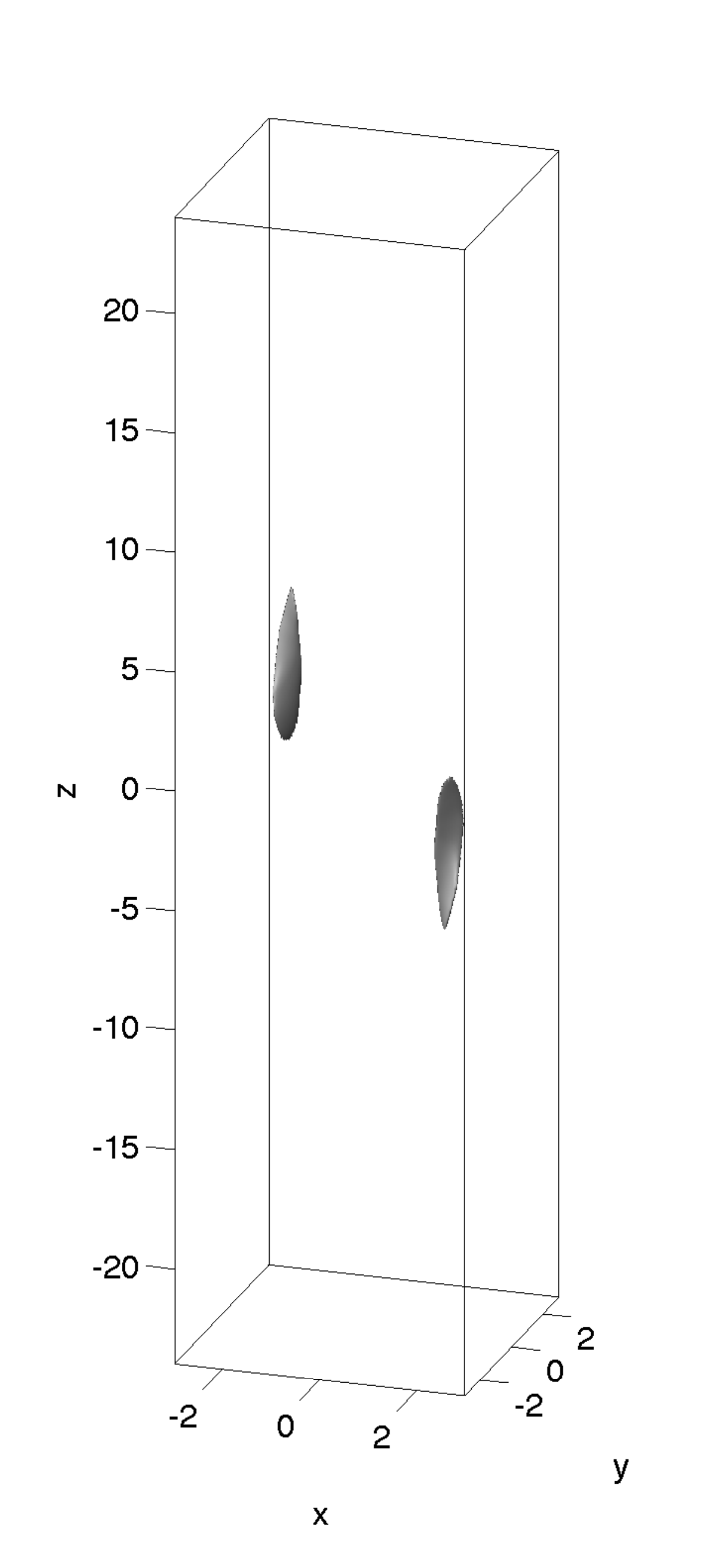}
\includegraphics[width=0.194\textwidth]{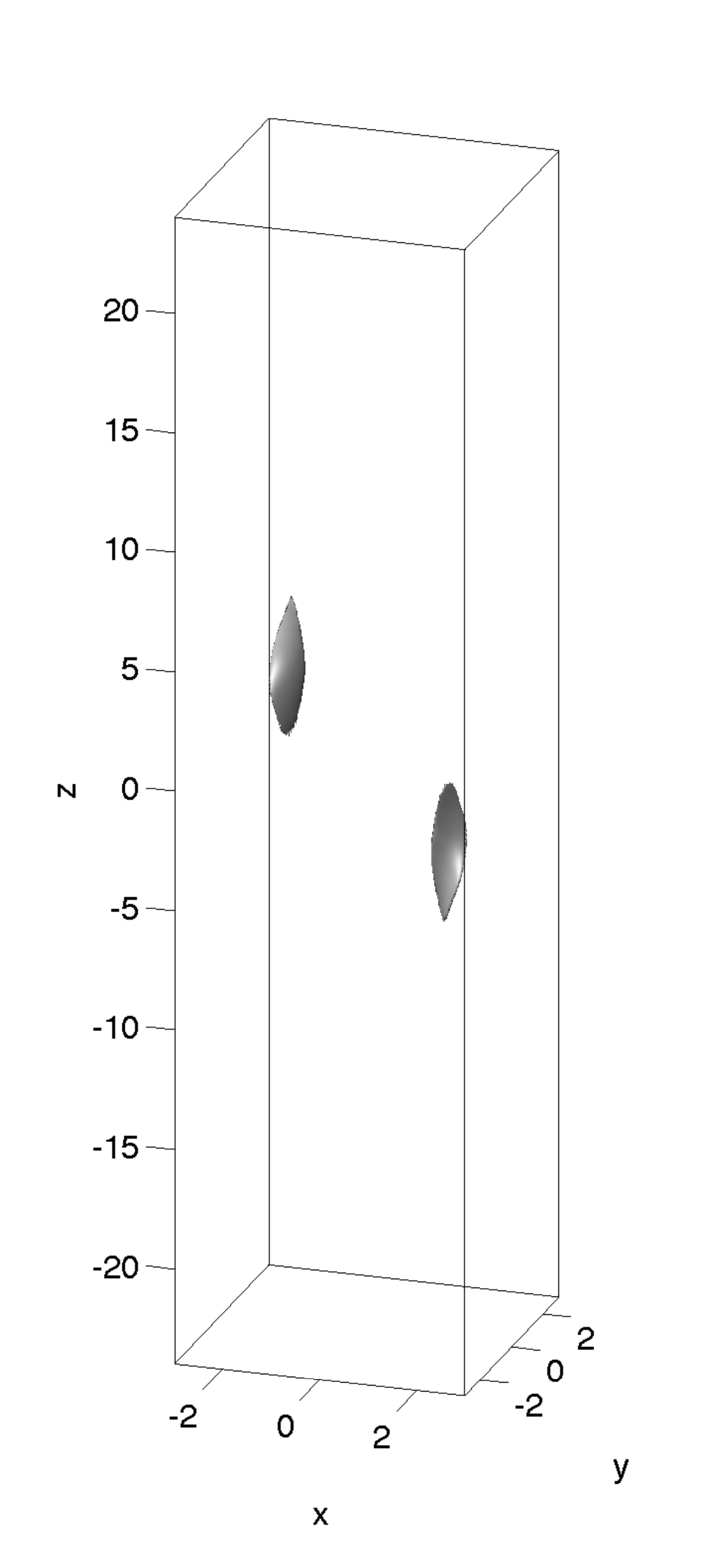}
\end{center}
\caption{Isosurfaces of current, $\vert \mathbf{J} \vert$,  at 50\% of the domain maximum
for a sequence of increasing times (from left to right, $t=4,6,8,10,12$) showing the
formation of the two initial current layers.}
\label{fig:j3D}
\end{figure*}

Figure~\ref{fig:j3D} shows isosurfaces of current at 50\% of the domain 
maximum ($\vert \mathbf{J} \vert =\vert \mathbf{J} \vert_{max}/2$) for a sequence of increasing
times (for the initial state see Fig.~\ref{fig:initialstate}).
In the early stages ($t \in [0,4]$) the current diffuses slightly while maintaining its large 
scales in the perpendicular direction. A symmetric evolution follows and after the phase of 
current growth two current concentrations 
are present, centered at $z=3.4$ and $z=-3.6$.  We call these the two `initial current layers'.

\begin{figure*}[]
\begin{center}
\includegraphics[width=0.405\textwidth]{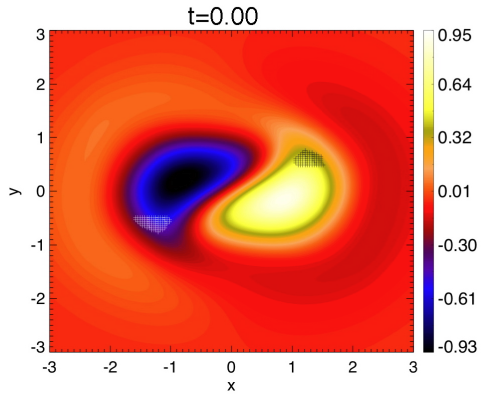}
\includegraphics[width=0.41\textwidth]{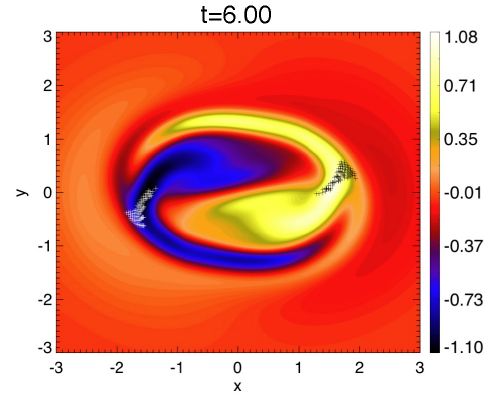}
\includegraphics[width=0.41\textwidth]{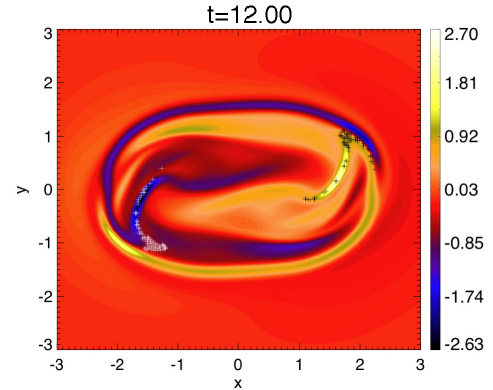}
\includegraphics[width=0.4\textwidth]{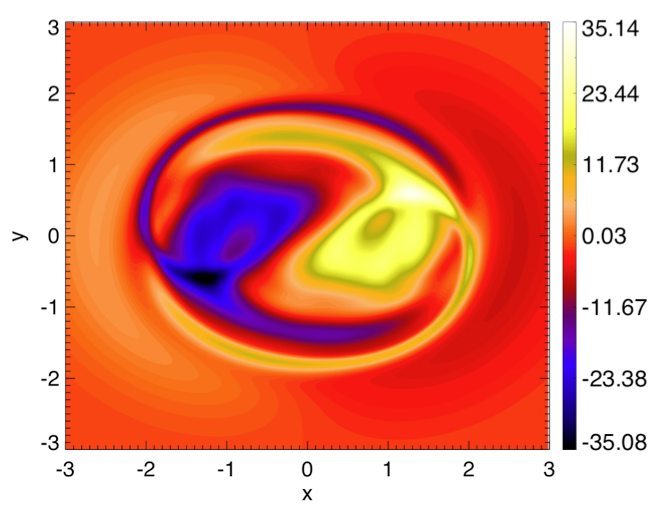}
\end{center}
\caption{Contours of the vertical component of the current,
$ {J}_{z}$, in the central plane, $z=0$, at increasing times, illustrating the formation 
of the two initial current layers (first three images).
The lower right-hand image shows contours of integrated parallel current along field 
lines in the initial state at the central plane $z=0$.
From this quantity field line tracers for the locations of initially
high integrated parallel current have been determined and marked in crosses in the
previous frames, as described in the main text.}
\label{fig:jevolution}
\end{figure*}

The formation of these two initial current layers is best illustrated by considering a 
horizontal cross section through the central plane ($z=0$).   Figure~\ref{fig:jevolution}
shows contours of the vertical component of current ($\vert {J}_{z} \vert$) in that plane
at $t=0, 6, 12$.
The $z$-component is taken since it significantly dominates over the two horizontal
components, as evident in the shape of the current sheets (Fig.~\ref{fig:j3D}).
Note that in order to incorporate both sheets the cross-sectional plane chosen, 
$z=0$, does not intersect the centre of either current sheet 
and so the magnitude of current in this plane is somewhat low in comparison 
to the domain maximum.
The collapse of the two oppositely signed large-scale fingers of current present in the initial state
into two thin current sheets of correspondingly the same sign is clearly shown.  
Also evident is the formation of  a weaker current envelope around the braided flux, 
separating it from the uniform background field. 
Cross-sections of $\vert {J}_{z} \vert$ in the horizontal planes through the centres of the
two current sheets are shown in the final two images of Fig.~\ref{fig:vel_buildup}.

\subsection{Predictors of Current Layers}

The field lines along which these two current sheets form turns out to be well 
predicted by the regions of  high integrated parallel current in the initial state.
In resistive MHD the integrated parallel current is related to the integrated parallel
electric field via the relation $\int J_{\|} dl = \eta \int E_{\|} dl$ (in the case of a uniform
resistivity $\eta$, as considered here).
The integrated parallel electric field is a key quantity for 3D magnetic reconnection;
for a localised reconnection region the maximum value of $\int E_{\|} dl$ determines
the rate of reconnection.

Shown in the lower right-hand panel of Fig.~\ref{fig:jevolution} 
 are contours of the integrated parallel current,
$\int J_{\|} dl$, in the initial state, $t=0$, where the path of the integral is taken over 
magnetic field lines.  Again the contour is shown in the central plane ($z=0$).  
To obtain this contour map, field lines have been integrated through
$160^{2}$ grid of points covering the domain $x,y \in [-3,3]^{2}$, $z=0$.
Two peaks in the quantity are present and the structure is quite different to that of 
the current itself in the initial state.
We now identify those field lines in this tracing procedure for which the value of 
$\vert \int J_{\|} dl \vert$ is greater than or equal to 75\% of the domain maximum,
noting the locations where they intersect the lower boundary 
(where the locations of the field lines are held fixed).   For the sequence of
 times of Fig.~\ref{fig:jevolution}  we trace field lines starting from those 
locations on the lower boundary up through the domain and mark with a cross 
in that same figure their point of  intersection with the $z=0$ plane.  
Here the difference in colours indicates field lines with positive (black crosses) and 
negative (white crosses) integrated parallel current, although this distinction is made only 
to facilitate  identification of the locations. 
It is found that these field lines, traced from the initial locations of high integrated parallel current,
are good indicators for the locations of formation of the two current layers. Since the flux on the lower
boundary is held fixed these may be identified as the same field lines for as long as the 
evolution remains ideal.  Whilst the evolution will be ideal only during the 
early stages of the simulation it is clear that the tracers do, nevertheless,
provide a useful predictor for the locations of current sheet formation.

\begin{figure}[]
\begin{center}
\includegraphics[width=0.43\textwidth]{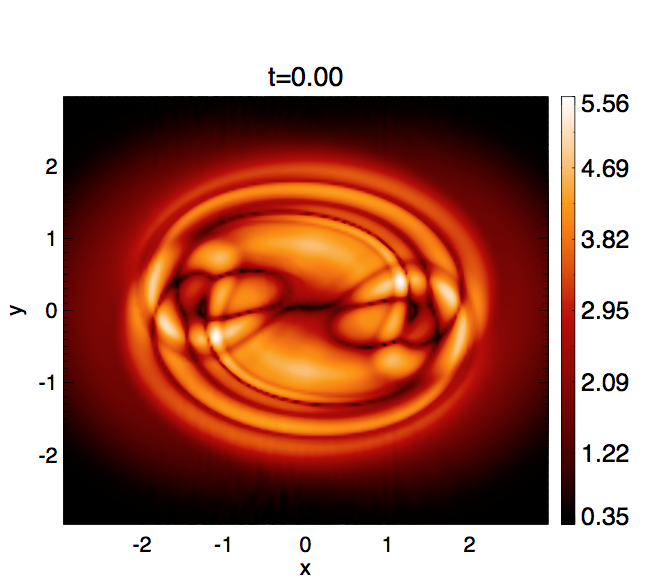}
\includegraphics[width=0.43\textwidth]{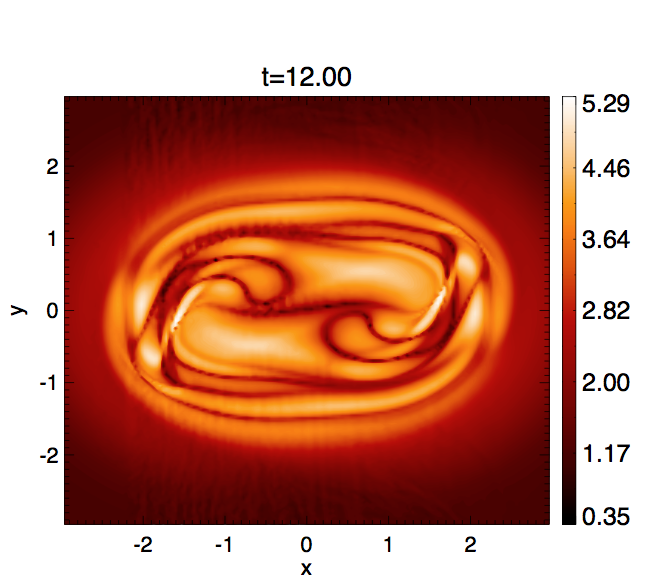}
\end{center}
\caption{Contours of the squashing factor, specifically $\log_{10}\left(Q\right)$,
in the central plane, $z=0$, for {\it(left)} the initial state, $t=0$ and {\it(right)} 
the time $t=12$.}
\label{fig:Qt0z0}
\end{figure}

Locations of high integrated parallel current 
are not a commonly used indicator for current sheet formation.    Indeed it is quasi-separatrix 
layers (QSLs), regions where the field-line connectivity varies strongly (Priest \& D{\'e}moulin, 1995) 
that are widely thought of as likely sites of current sheet formation.
To identify QSLs (as well as their intersections, hyperbolic flux tubes or HFTs) the squashing
factor  (Titov {\it et al.}~2002) is used.  Usually denoted by $Q$, the squashing factor is an
indictor of field line connectivity and takes on high values in regions where the field line mapping
is strongly distorted.  Regions of high $Q$ outline QSLs.  As discussed in 
Wilmot-Smith {\it et al.}~(2009b), the braided magnetic field taken as the initial state here contains 
several QSLs.
Contours of the squashing factor, $Q$, in the central plane ($z=0$) are shown in
Fig.~\ref{fig:Qt0z0} at $t=0$ (left) and $t=12$ (right).  For the calculation again $160^{2}$ points over 
the region  $x,y \in [-3,3]^{2}$ have been used, a number comparable to the grid resolution.

Several regions of high $Q$ are present in both snapshots.  
The two regions of highest $Q$ in the initial state are associated with the two initial 
current layers at the later time, that is, the current sheets have formed along QSLs of the field.
At the same time, several other regions of high $Q$ are present (both at $t=0$ and
$t=12$) that are not associated with any particular current features.  
For example, in the initial state there are eight distinct regions where $\log_{10}\left(Q\right)$
is greater than 85\% of its maximum value but
 only two of these regions correspond to particular current features at $t=12$.
These results suggest that the integrated parallel electric current and the squashing
factor could be used in conjunction for predicting current sheet formation more accurately
than by using $Q$ alone.

\subsection{Plasma Flows and Reconnection}

\begin{figure*}[]
\begin{center}
\includegraphics[width=0.4\textwidth]{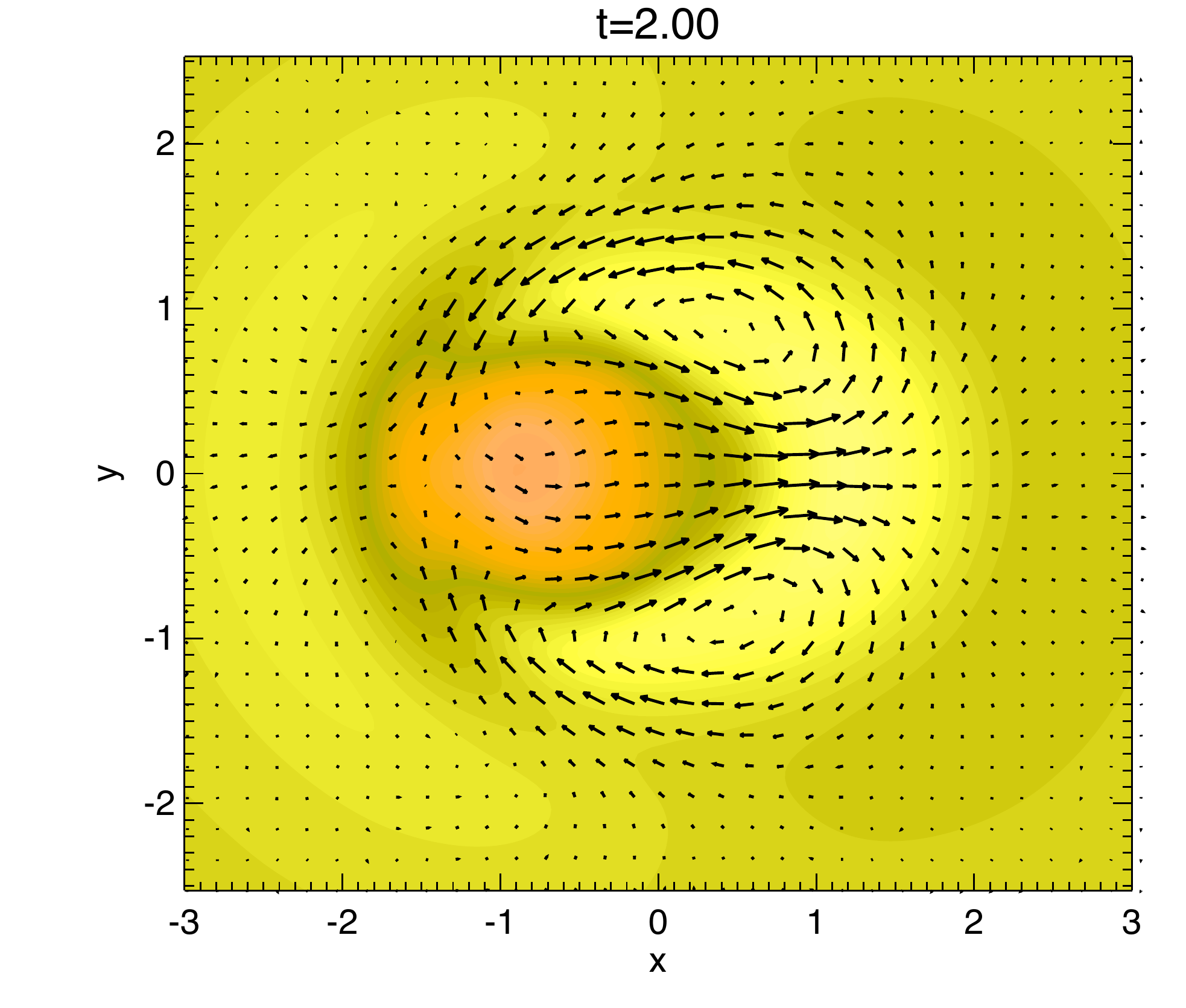}
\includegraphics[width=0.4\textwidth]{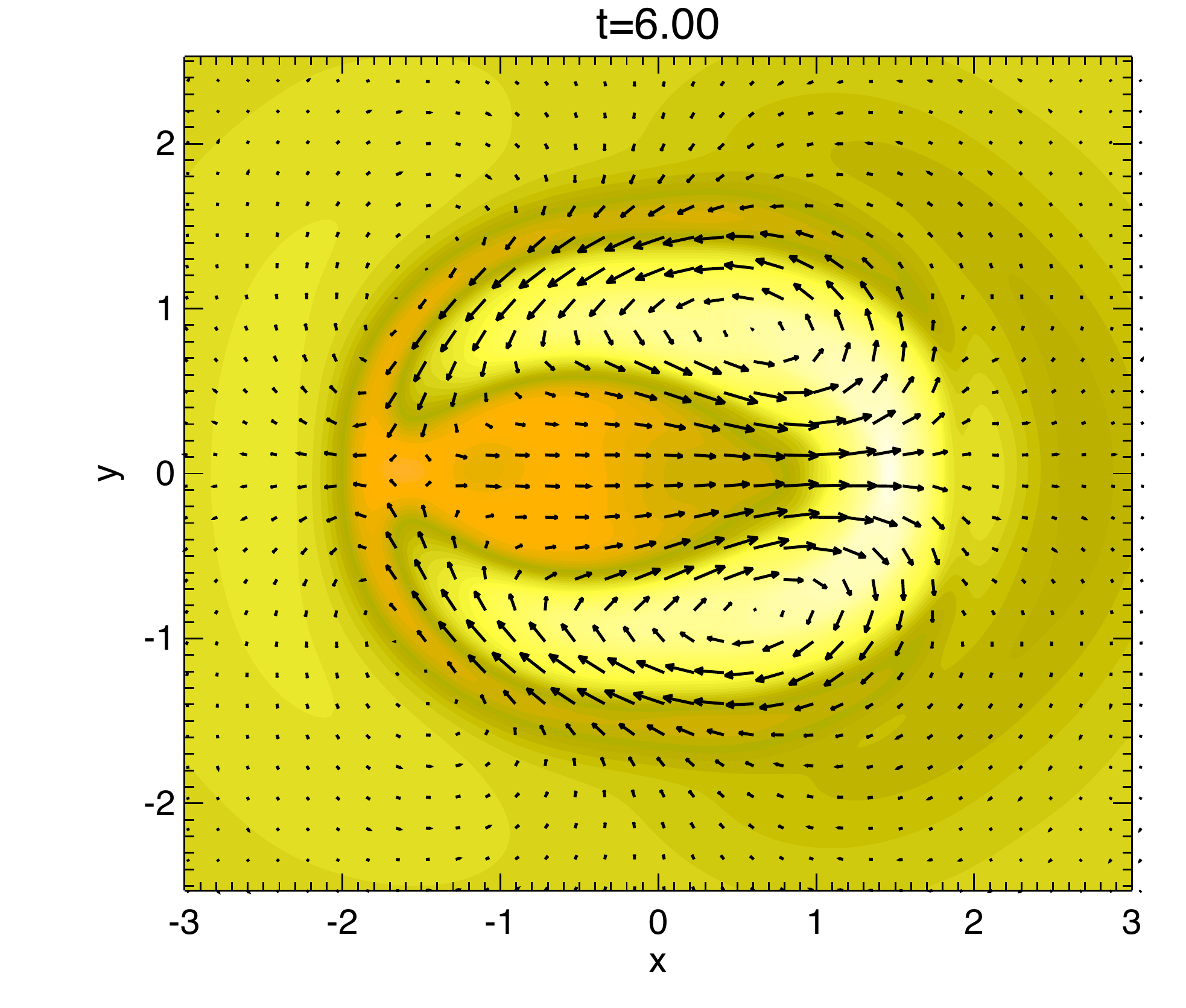}
\includegraphics[width=0.4\textwidth]{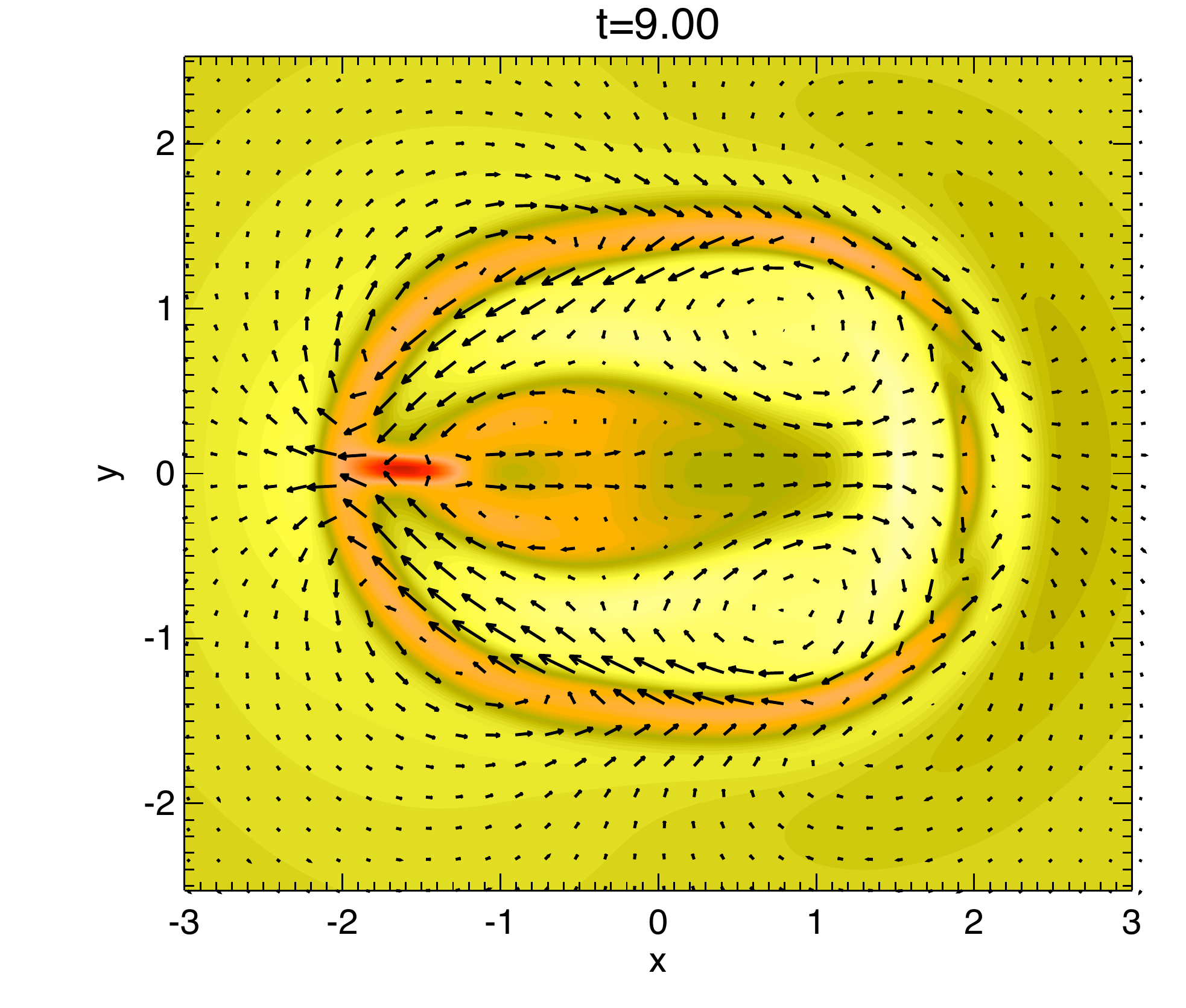}
\includegraphics[width=0.4\textwidth]{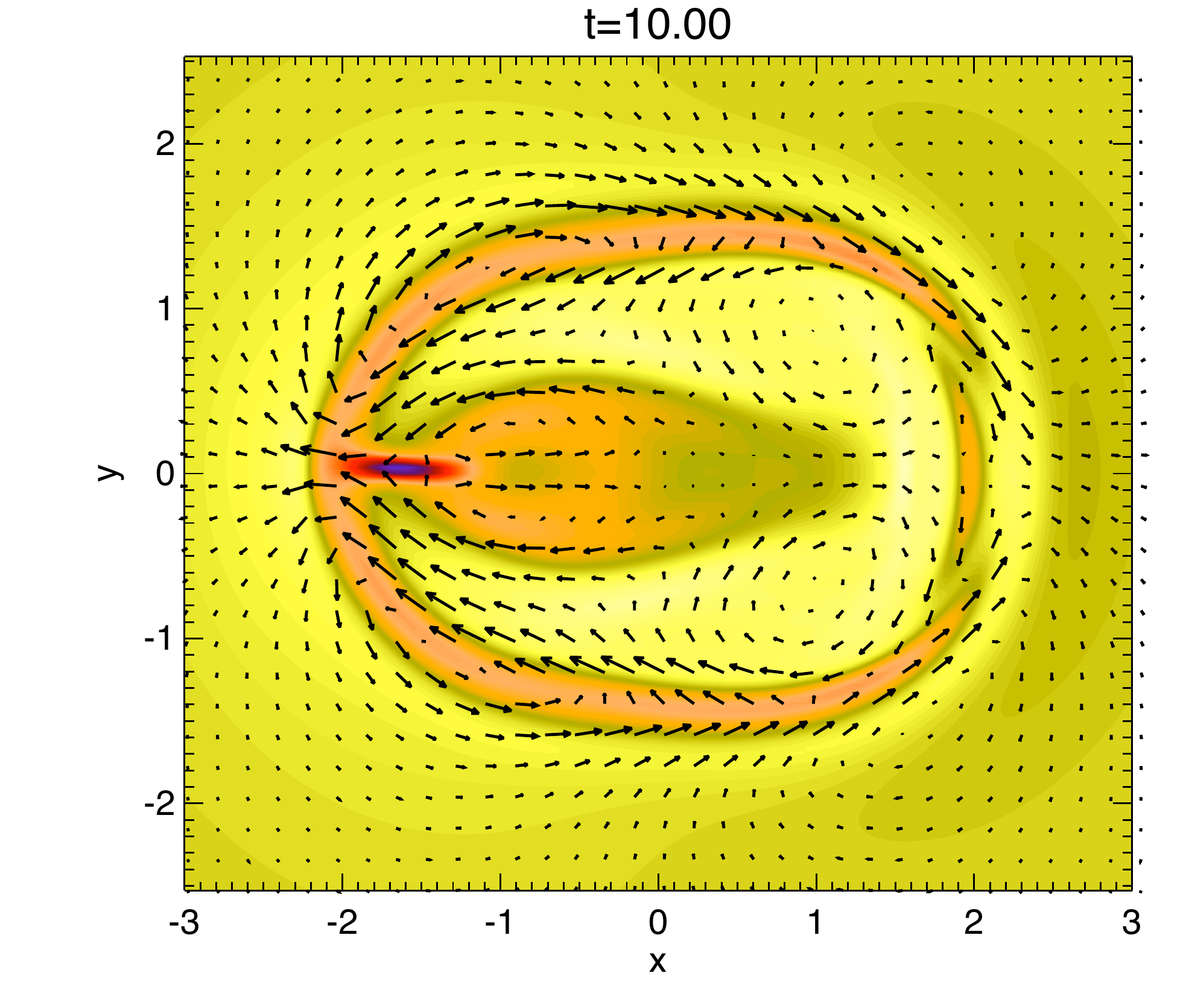}
\includegraphics[width=0.4\textwidth]{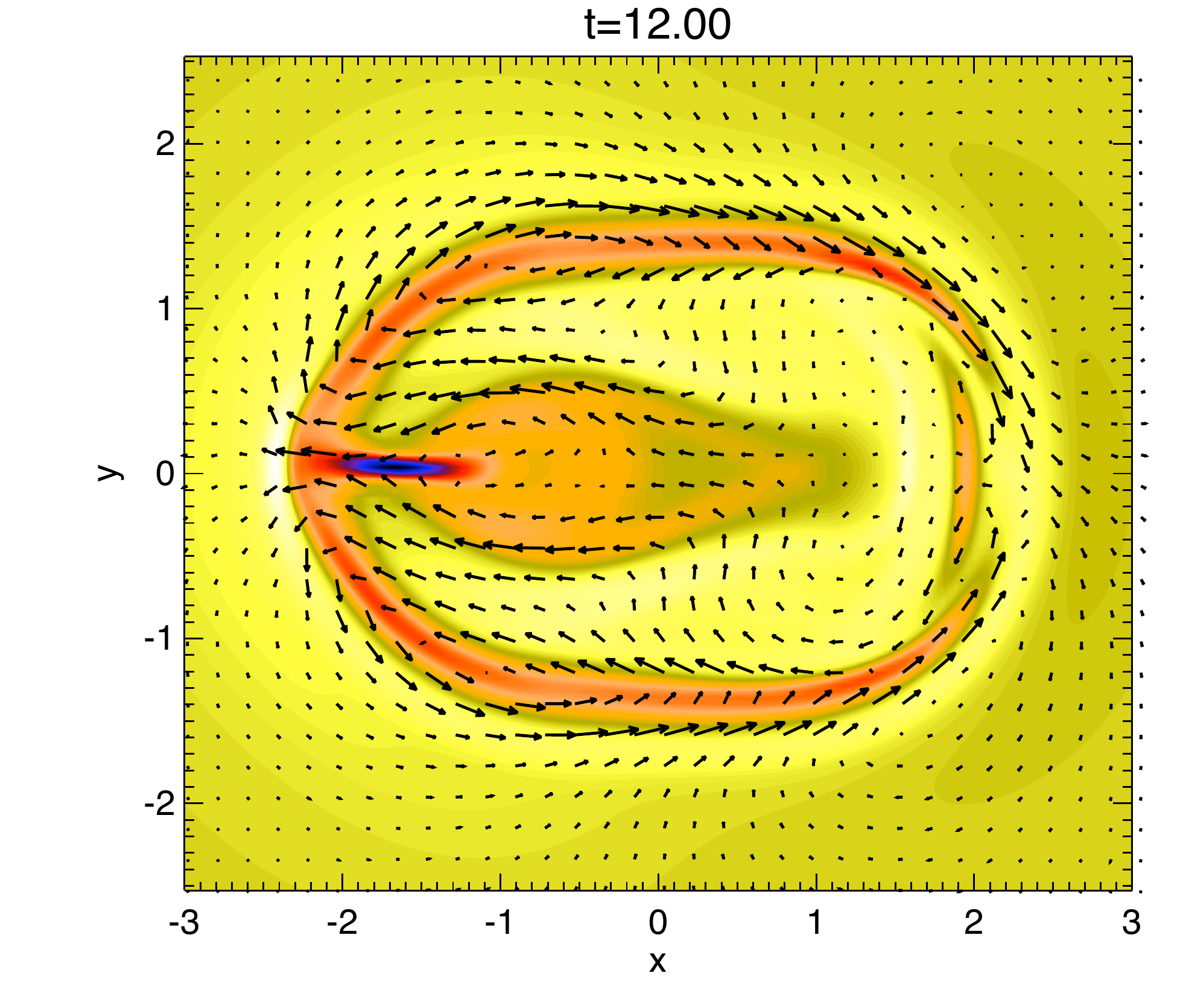}
\includegraphics[width=0.063\textwidth]{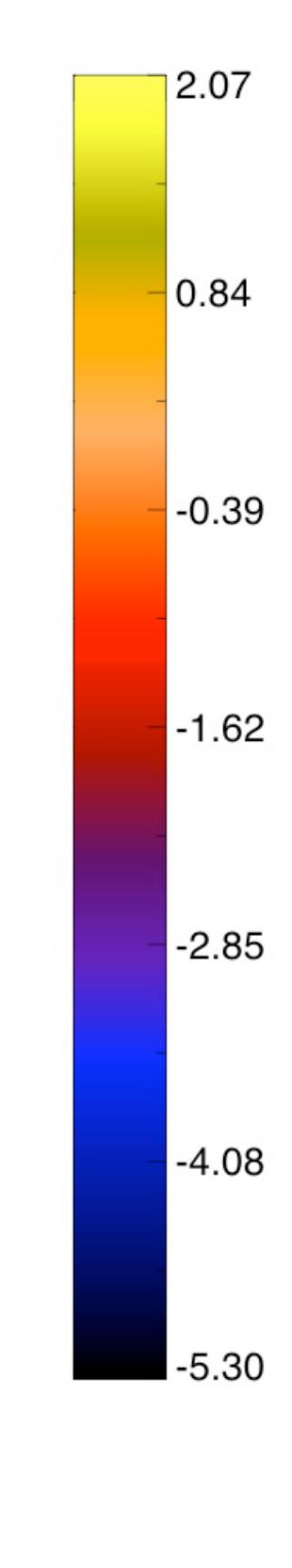}
\includegraphics[width=0.4\textwidth]{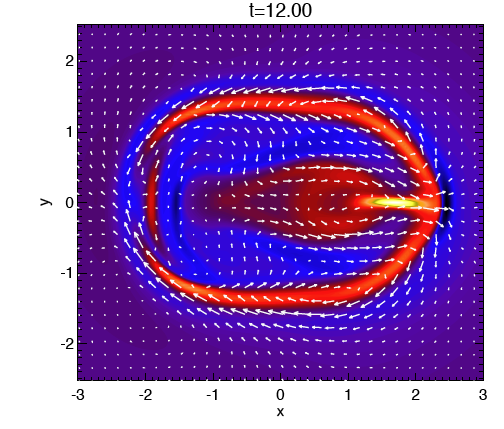}
\includegraphics[width=0.0585\textwidth]{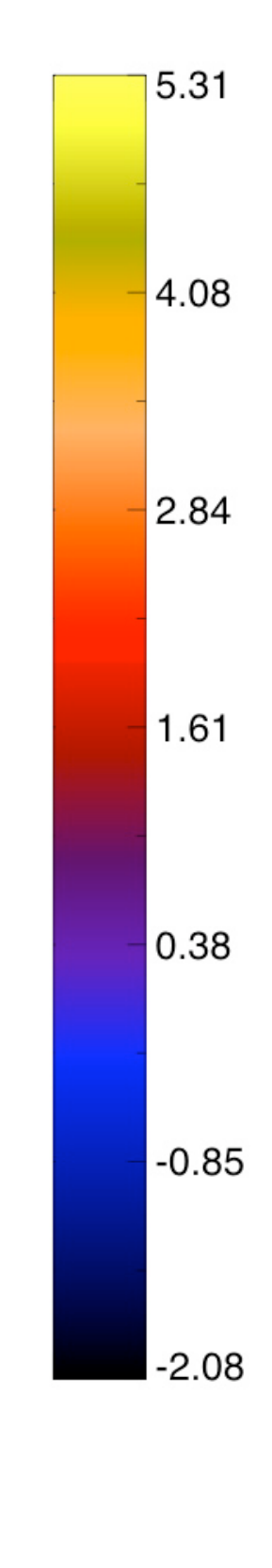}
\end{center}
\caption{Arrows indicate plasma velocities $[V_{x},V_{y}]$ in horizontal cross-sectional planes,  
superimposed on the vertical component of current, $J_{z}$.  The first five images
show structures about the upper current sheet (taking the plane $z=3.4$) while the final
image is for the lower current sheet (taking the plane $z=-3.6$).  The images are given
at various times, as indicated in the figures.}
\label{fig:vel_buildup}
\end{figure*}

We move now to consider the nature of the plasma flows within the domain.
First recall that the low values of the plasma beta (Sec.~\ref{sec:numerical}) 
imply the dynamics will be dominated by the Lorentz force rather than the gas pressure.
With both the magnetic field and the current having stronger vertical than horizontal
components we have that the Lorentz force is primarily in the horizontal direction
and so, similarly, are the plasma flows.
The Lorentz force drives a flow with a dipolar structure in the six regions of initially
strongest current with direction dependent
on the sign of the twist in that region.  An illustration of such a flow is shown 
for the plane $z=3.4$ (a negative twist region corresponding to one of the sites of current
sheet formation) in the first (upper left) image of Fig.~\ref{fig:vel_buildup}.

The next four images in that figure show the development of the flow for a sequence
of increasing times up to the point of maximum current.  In the sequence the length of the
arrows indicating flow direction have been normalized to each image.  The background
contours show the vertical component of current in that plane with the colour scale normalised
to the current at $t=12$ (lower--left-hand image), given by the first colour-bar.  
The sequence clearly shows the association of the stagnation part of the dipolar flow with the 
current intensification.
 The out-flowing plasma from the location of stagnation sets up a counter-flow to 
the initial dipolar structure leading to oppositely directed flows on either side of the weak 
enclosing current sheath.  The final image shows plasma flows and current in the plane $z=-3.6$ 
at $t=12$.  This  cross-section is across the second current sheet and shows the naturally 
expected inversion of the flow direction.
The result is that the global flow structure is dominated by rotational components the direction
of which varies both vertically along the structure and on either side ($y>0$, $y<0$) of the braided
field.

As already indicated by Fig.~\ref{fig:jmax}, at $t=12$ the maximum magnitude of the plasma 
flows ($\vert \mathbf{V} \vert_{max} \sim 0.24$) is a significant fraction of the Alfv{\'e}n speed
(${V}_{A} \sim  1.2$). These strongest flows are associated with the 
global rotation and not with outflows from the two initial current layers (which have an 
associated $\vert \mathbf{V} \vert_{max} \sim 0.15$).  However, we do expect magnetic reconnection 
to be taking place across these current sheets and so proceed to consider the nature of this
reconnection.  For this we concentrate again on the initial current layer centered at $z=3.4$ 
(a similar situation occurs about the other current layer)
{ and focus on the structure of the field and flows in the plane perpendicular to the
magnetic field at the location of maximum current magnitude,  $\vert J\vert$.}

\begin{figure*}[]
\begin{center}
\includegraphics[width=0.4\textwidth]{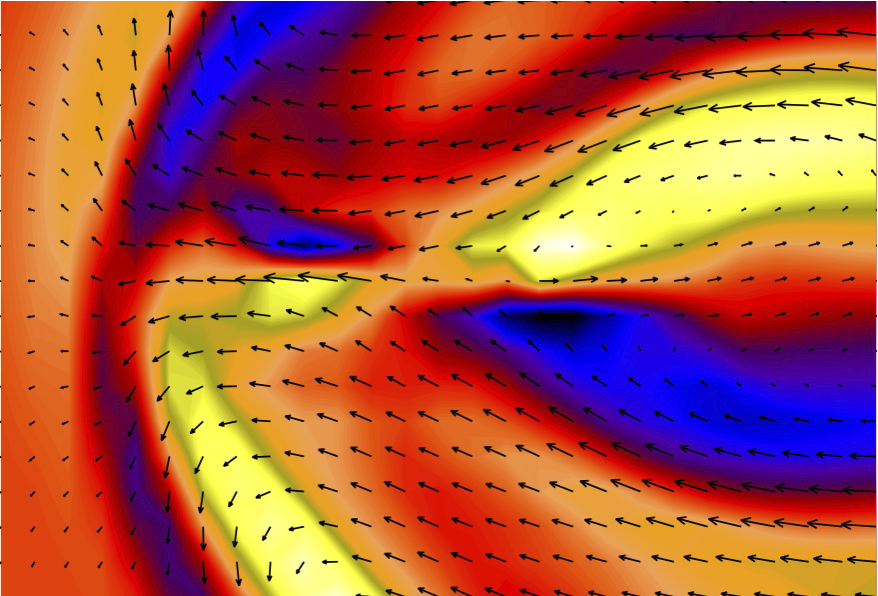} 
\includegraphics[width=0.395\textwidth]{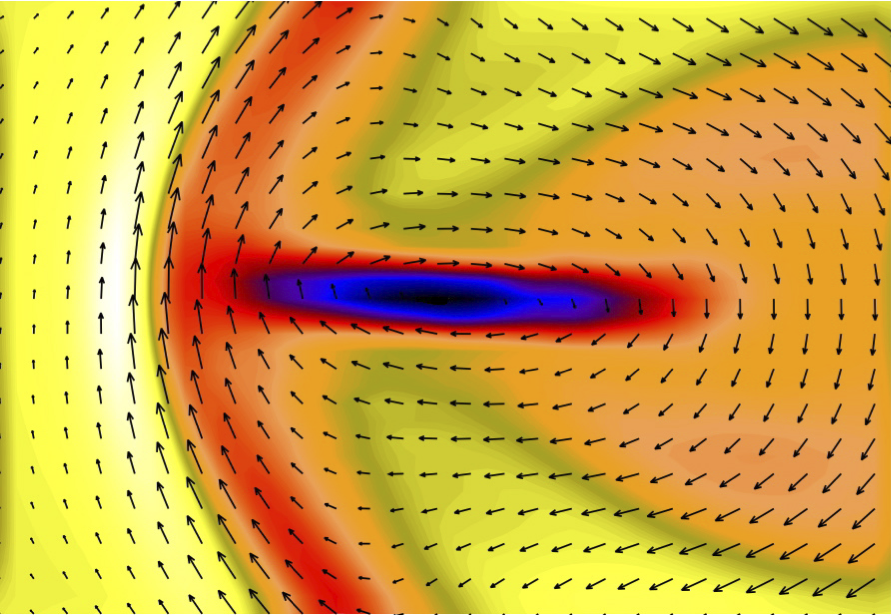}
\end{center}
\caption{{ Here various quantities are considered in the plane perpendicular to the magnetic
field at the region of maximum (negative) current.
({\it Left})  Arrows indicate plasma flows in that plane while the coloured contours show the out-of-plane
component of vorticity.
({\it Right}) Arrows indicate magnetic field components in the plane superimposed
on contours of the out-of-plane current.
The colour table for each is {\it blue--yellow} for {\it negative -- positive}.
}}
\label{fig:vort}
\end{figure*}

In the left-hand image of Fig.~\ref{fig:vort} we indicate in more detail 
the nature of the stagnation flow about the current sheet.  
{ Superimposed on the background are contours of the out-of plane component of vorticity, 
i.e. the component of vorticity in the direction of the magnetic field at the location of maximum current.
The vorticity shows a quadrupolar configuration around the current sheet.

Perhaps more of a surprise  is the structure of the magnetic field in the region.
The right-hand image of  Fig.~\ref{fig:vort} shows the components of the magnetic 
field in the cross-sectional plane under consideration.    The magnetic field
is shown to have an elliptic configuration about the current sheet, i.e.~about the 
reconnection region.  
This finding contrasts with the two-dimensional picture of magnetic reconnection}
under which the process can only occur at a hyperbolic (X-type) null-point of the magnetic field.
In three-dimensions a much wider variety of possible reconnection sites exist.
Reconnection may be associated with  3D null-points (e.g.~Lau \& Finn, 1990;  Priest \& Pontin, 2009),
magnetic separators (which connect two null-points, e.g.~Longcope \& Cowley, 1996; Pontin \&
Craig, 2006; Haynes {\it et al.}, 2007), or may occur in the absence of any such topological 
features (Schindler {\it et al.}, 1988), the latter sometimes termed `non-null reconnection'.
In particular, the local magnetic field structure need not be hyperbolic but may be elliptic
(Hornig \& Priest 2003), as recently found in some 3D numerical simulations of reconnection.
For example, Wilmot-Smith \& De Moortel (2007)  considered reconnection occurring along a 
quasi-separator and found an elliptic field structure in perpendicular cross-sections.
The separator configuration of Parnell {\it et al.}~(2010) showed an elliptic 
structure along a significant length of the separator under consideration.
 Parnell {\it et al.}~(2010) also discussed the separator case theoretically, concluding an elliptic
 configuration would be a generic situation given a sufficiently strong current density
 along the separator.
Our findings demonstrate that an elliptic field configuration may be present about reconnection 
sites in the non-null case.
Additionally, tracking these field lines back to the initial setup, an elliptic perpendicular
field configuration is again found indicating that locally hyperbolic structures are not necessary for 
current intensification.
As previously discussed, the squashing factor ($Q$) at the two reconnection sites is very 
high which demonstrates a further point; 
regions of highest-$Q$ within a domain may have a locally elliptic field configuration.

\section{Nature of Instability}
\label{sec:nature}

Evidently the initial magnetic field configuration is not in a stable equilibrium.
Since an exact equilibrium of the ideal relaxation code employed in Wilmot-Smith {\it et al.}~(2009a)
is known to be linearly ideally stable, the {  lack of stability could arise from one of a number of factors:}
\begin{enumerate}

\item A resistive instability.    The relaxed state of Wilmot-Smith {\it et al.}~(2009a) contains small scales in 
the integrated parallel current.  Small enough scales (for a given resistivity) in this quantity are incompatible 
with a resistive equilibrium.

\item {  A non-linear} ideal instability not previously found in the ideal evolution of Wilmot-Smith {\it et al.}~(2009a) 
{  since that evolution only guaranteed linear stability}.

\item{   A lack of equilibrium in the initial state} either since:
\begin{itemize}

\item The path to equilibrium of the ideal relaxation is fictitious and an exact equilibrium had not been reached.  
Numerical difficulties (described in Pontin {\it et al.}~2009) result in the final state of the ideal relaxation having 
$\vert {\bf J} \times {\bf B} \vert_{max} \approx 2 \times  10^{-2}$, i.e. is not perfectly force-free.
Thus the final state of Wilmot-Smith {\it et al.}~2009a need not be stable (or, indeed accessible via a real MHD 
relaxation dynamics).

\item  The technique used to transfer the initial state between codes has perturbed the magnetic 
field  {  further from force-free.}

\end{itemize}
\end{enumerate}

At the beginning of Sec.~\ref{sec:results} we {  gave one suggestion, that the formation of the current layers 
may be due to an ideal instability as evidenced by the lack of dependence of kinetic energy
on resistivity (see Fig.~\ref{fig:jmax}).  Here we seek to determine additional information that may identify
the cause of the dynamical evolution.
We have noted that the integrated parallel current is a good predictor of the location of the 
two initial current layers. The integrated parallel current is a global quantity and so the global structure of the field may
play a key role. The global structure is also important in,}
for example, the kink instability where a magnetic field with a set number of turns per unit length 
becomes unstable as more turns are added by increasing the length of the system.  
Whilst the kink-instability as it is usually considered applies to a tube with a well-defined single
axis a similar kink-like instability, dependent on the total twist of the system, {  may} apply
to our braided field.

{  With these considerations in mind we examine the evolution of only} the middle section of the 
initial state of the braided magnetic field that is, we cut-out the section of the field in the above described
experiment that lies in $z \in [-8,8]$, $x,y \in [-6,6]$ at $t=0$.  This field is inserted as an initial 
condition in a new run, now keeping the flux fixed on $z=\pm 8$, the new upper and lower boundaries
of the domain.  To maintain consistency we use the same resolution in the horizontal direction $320^{2}$ and a 
similar effective resolution in the vertical direction, $128$ cells over $z \in [-8,8]$.  

\begin{figure}[]
\begin{center}
\includegraphics[width=0.51\textwidth]{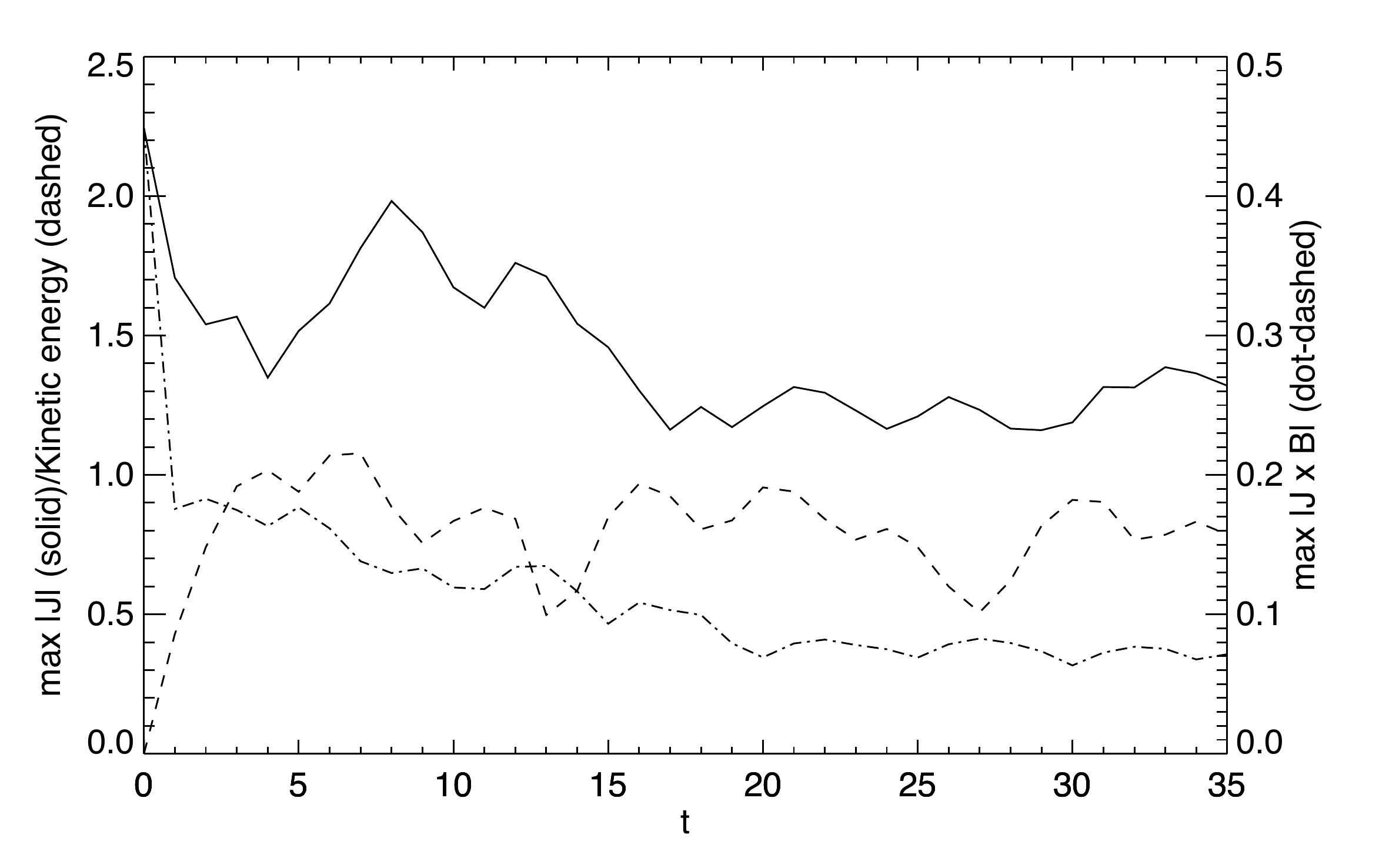}
\includegraphics[width=0.4\textwidth]{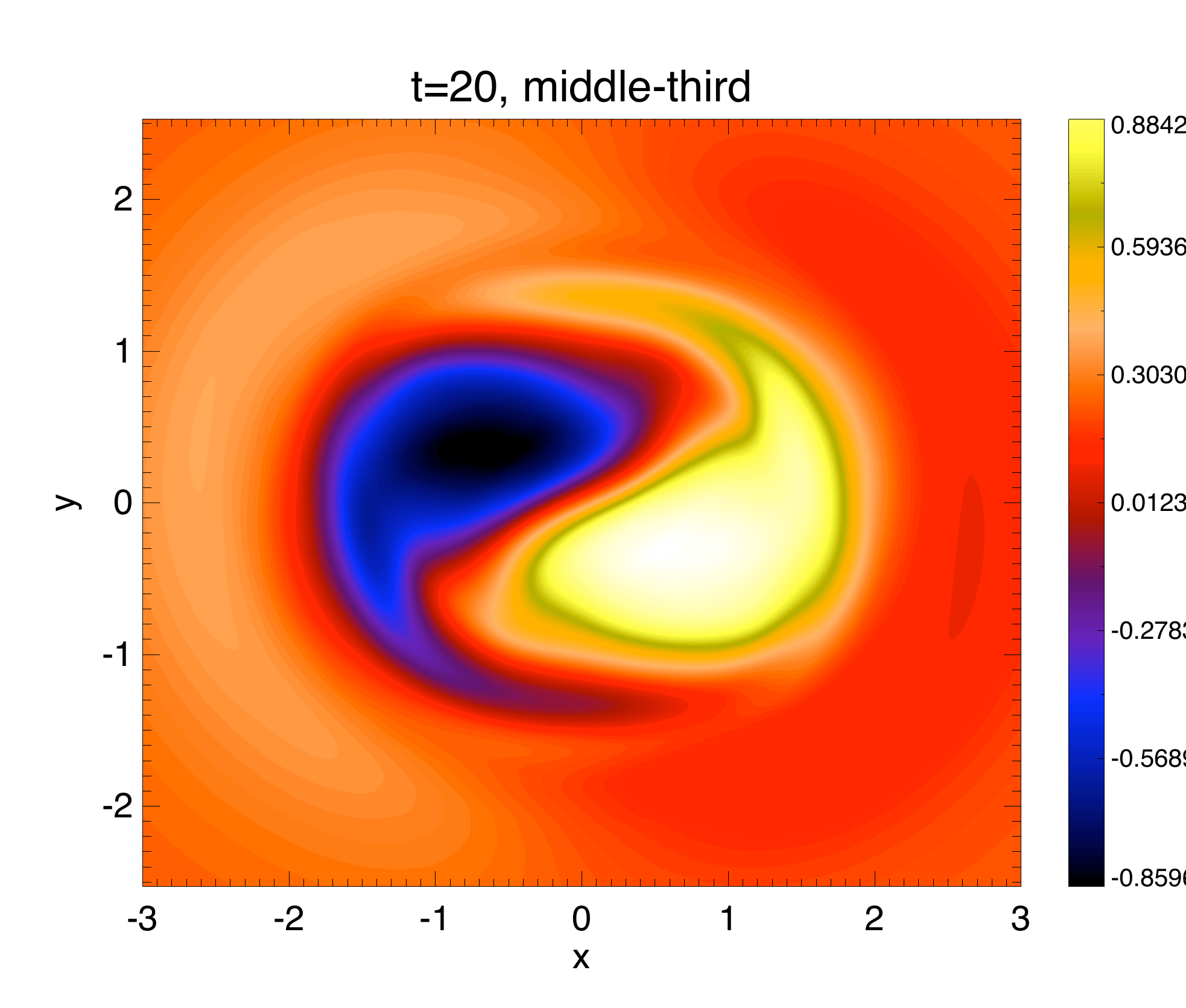}
\end{center}
\caption{({\it Left}) Vertical component of current in the central plane ($z=0$) at $t=20$ for the
an MHD evolution on only the central section, $z\in[-8.0,8.0]$ of  the full field, as 
described in the main text. ({\it Right}) Maximum current $\vert {\bf J} \vert$ (solid line) 
and total kinetic energy (dashed line)  {  and maximum Lorentz force (dot-dashed line)}
over time for the same field.}
\label{fig:middlethird}
\end{figure}

In the evolution of this new `middle-third' field we find the system adjusts from its initial condition (with zero plasma velocity) 
to an approximate equilibrium 
{  in which the current structure is qualitatively similar to that of the initial state.}
To illustrate, contours of current in the central plane ($z=0$) are 
shown at $t=20$  in Fig.~\ref{fig:middlethird} (right) which may be compared with  Fig.~\ref{fig:jevolution} 
(upper left) where the corresponding contours in the initial state, $t=0$, are shown.  
The maximum current in the domain is shown as a solid line in {  the left-hand image of Fig.~\ref{fig:middlethird} 
(left-hand axis), the kinetic energy integrated over $x,y \in [-3,3]$, $z \in [-8,8]$ as a dashed line
(left-hand axis) and the maximum Lorentz force within the domain as a dot-dashed line
(right-hand axis).
As in the evolution of the full field, the  Lorentz force in this initial state is  high as a result of the transfer 
between grids and decreases rapidly over the first few time units.  
However, the Lorentz force subsequently stabilises at a low value as the system readjusts to equilibrium.
The result that the middle section of the braid alone is in an equilibrium state 
suggests that an instability is present in the full braided field (rather than a lack of equilibrium 
due to numerical artefacts).  Furthermore the instability is of a long-wavelength type 
with the full structure of the braided field being key.}

{  Returning to the evidence of Figure~\ref{fig:jmax},
during the time $t \in [8,12]$ where the instability  is clear in the current growth}
there is only a very slight dependence of the maximum velocity within the domain on resistivity 
and no increase in kinetic energy.  This effect may be due to the confining nature of the strong background 
field external to the braided field structure which results in a deflection of the outflowing plasma around the boundary
of the braided field (see Fig.~\ref{fig:vel_buildup}). Prior to $t \approx 8$ no dependence of the flow on 
resistivity is seen suggesting an ideal dynamics where the system is adjusting to the distance from equilibrium.
{   The current growth in $t \in [8,12]$ does have a clear separation according to resistivity $\eta$. 
However the increase is linear (rather than exponential) suggesting a dominant non-linear phase and 
the growth is slower for higher resistivity suggesting the non-linear  phase is damped by resistive effects.}

\section{Conclusions}
\label{sec:discussion}

A previous paper (Wilmot-Smith {\it et al.}~2009a) considered the ideal relaxation of a 
braided magnetic field towards a force-free equilibrium.  Here we have taken the final state of that 
relaxation and used it as an initial condition for a full resistive MHD simulation.  

The braided field is not in a stable equilibrium; two thin current sheets form after a short time
(around a quarter of the time for an Alfv{\'e}n wave to cross the numerical box in the vertical direction).
The linear rate of increase of current density and the maximum strength of the current is found to increase 
with decreasing resistivity although the evolution of the total kinetic energy in the domain is independent 
of resolution.  We conclude that the instability is {  possibly} an ideal one although the details
of how it occurs have not been determined.
The wavelength of the instability was tested by considering the MHD evolution of the middle
third section of the braid alone and this new field was found to be stable.  Hence
a long-wavelength dependence is implied.

The initial configuration contains many regions of high squashing factor, $Q$, corresponding to
quasi-separatrix layers.  Plasma flow across QSLs is often thought likely to lead to current 
sheet formation.  The two current sheets that form here do align with two of the highest regions of 
$Q$ although the remaining regions of high $Q$ do not correspond to any particular current features.
The locations of the two current layers are well predicted by the 
peak values of the integrated  parallel current in the initial state.  In the central plane this quantity
shows two clear peaks and it is at these regions that the two current layers later form.

These two current layers correspond to reconnection sites.   Perpendicular to the layers the magnetic 
field has an elliptic structure, 
an admissible property of three dimensional reconnection that has only recently been found.
The flow about the reconnection sites is of an asymmetric stagnation type, although large-scale rotational flows dominate the global structure; the effect of  reconnection is not a strong acceleration of the flow but a more subtle 
 untwisting process leading to the change in magnetic field topology.

The longer-term evolution of the system will be considered in a future paper.

\vspace*{0.1cm}

\begin{appendix}
\noindent
{\large {\bf Appendix}} 
\vspace*{0.1cm}

\noindent
Equation~(\ref{eq:Atoexplain})
 can be derived from an evolution for the vector potential of a frozen-in magnetic field:
\begin{equation}
	\frac{\partial {\bf A}}{\partial t} + \nabla  \left( {\bf V \!  \cdot  \! A}\right)
	- {\bf V} \times \nabla \times {\bf A} = 0.   \label{vectorpotvolu}
\end{equation}
This equation is equivalent to the Lie-derivative for a differential one-form, $\alpha = A_i dx^i$, associated with
 the vector field ${\bf A}$ (Hornig, 1997). Hence (\ref{vectorpotvolu}) can be written as 
\begin{equation}
	\frac{\partial \alpha}{\partial t} + L_{\bf v} \alpha = 0, 
\end{equation}
and we can express $\alpha(t=0) = F^* \alpha(t)$ (Abraham {\it et al.}~1988), 
or more conveniently $\alpha(t) = (F^{-1})^* \alpha(t=0)$,
where $F: X \longrightarrow x$ maps the initial to the final coordinates and the star indicates 
the pull-back operation. This last equation written in components of the vector potential is (\ref {eq:Atoexplain}).
\end{appendix}

\vspace*{0.3cm}

\noindent
{\bf Acknowledgements} \\
\noindent
The authors would like to thank Nasser Al-Salti for helpful suggestions and Klaus Galsgaard 
for the use of his code.
A.W.S. and G.H. acknowledge financial support from the UK's STFC. 
The simulations described were carried out on the STFC and SFC (SRIF) funded
linux clusters of the UKMHD consortium at the University of St Andrews.

{
The Transition Region and Coronal Explorer, TRACE, is a mission of the Stanford-Lockheed 
Institute for Space Research (a joint program of the Lockheed-Martin Advanced Technology 
Centre's Solar and Astrophysics Laboratory and Stanford's Solar Observatories Group), and 
part of the NASA Small Explorer program.}

\newpage
\noindent
{\large {\bf References}}
\vspace*{0.2cm}

\noindent
Abraham, R., Marsden, J.E.,  Ratiu, T. 1988, Manifolds
Tensor Analysis and Applications, Applied Mathematical Sciences {  75},  370, Springer-Verlag, New York.

\noindent
Berger, M.A., Asgari-Targhi, M., 2009, ApJ, 705 347-355.

\noindent
Craig, I. J. D., Sneyd, A. D. 1986, {  ApJ}, {  311} 451-459.  

\noindent
Craig, I. J. D., Sneyd, A. D. 2005 {  Sol. Phys.}, {  232} 41-62.

\noindent
Galsgaard, K., Titov, V.S., Neukirch, T. 2003, {  ApJ}, {  595} 505.

\noindent 
Haynes, A.L., Parnell, C.E., Galsgaard, K., Priest, E.R. 2007, {  Proc. Roy. Soc. A}, {  463}
1097-1115.

\noindent
Hornig, G.  1997, Phys. Plasmas {  4}, 646-656.

\noindent
Hornig, G., Priest, E.R. 2003, {  Phys. Plasmas}, {  10}, 2712-2721.

\noindent
Janse, {\AA}. M., Low, B.C. 2009, {  ApJ}, {  690} 1089-1104.

\noindent 
Lau, Y-T.,  Finn, J.M. 1990, {  ApJ}, {  350} 672-691.

\noindent 
Longcope, D.W, Cowley, S.C. 1996, {  Phys. Plasmas}, {  3} (8) 2885-2897.

\noindent
Longcope, D., Strauss, H.R. 1994, {  ApJ}, {  437}, 851-859.

\noindent
Ng, C.S., Bhattacharjee, A. 1998, {  Phys. Plasmas}, {  5}, 4028-4040

\noindent
Nordlund, A., Galsgaard, K. 1997, A 3D MHD code for parallel computers. Technical
report, Astronomical Observatory, Copenhagen University.

\noindent
Parker, E.N. Spontaneous Current Sheets in Magnetic Fields With Applications
to Stellar X-Rays. 1994, Oxford University Press, Inc., New York (pp. 23-24).

\noindent
Parnell, C.E., Haynes, A.L., Galsgaard, K. 2010, {  JGR}, {115} A02102.

\noindent 
Pontin, D.I., Craig, I.J.D. 2006, {  ApJ}, {  642} 568-578.

\noindent
Pontin, D.I., Hornig, G., Wilmot-Smith, A.L., Craig, I.J.D. 2009, {  ApJ}, {  700} (2) 1449-1455.

\noindent
Priest, E.R., Demoulin, P. 1995, {  JGR}, {  100} (A12) 23443-23464.

\noindent 
Priest, E.R., Pontin, D.I. 2009, {  Phys. Plasmas}, {  16} 122101.

\noindent 
Schindler, K., Hesse, M., Birn, J. 1988, {  JGR}, {  93} 5547-5557.

\noindent
Titov, V.S., Hornig, G., D{\'e}moulin, P. 2002, {  JGR}, (A8) SSH 3-1 1164.

\noindent
van Ballegooijen, A.A. 1985, {  ApJ}, {  298}, 421.

\noindent 
Wilmot-Smith, A.L., De Moortel, I. 2007, {  A\&A}, {  473} 615-623.

\noindent
Wilmot-Smith, A.L., Hornig, G., Pontin, D. I. 2009a, {  ApJ}, {  696} 1339-1347.

\noindent
Wilmot-Smith, A.L., Hornig, G., Pontin, D.I. 2009, {  ApJ}, {  704} (2) 1288-1295.

\end{document}